\documentclass[article,twocolumn]{revtex4-2} 
\usepackage{blindtext}
\usepackage{titlesec} 
\usepackage{bm}
\usepackage{float}

\usepackage{graphicx} 
\usepackage{natbib} 
\usepackage{hyperref}
\usepackage[usenames,dvipsnames]{color} 
\usepackage{amsmath} 
\usepackage{amssymb} 
\usepackage{ifthen} 
\usepackage{tikz}
\usetikzlibrary{decorations.pathmorphing,decorations.markings}
\usepackage{makeidx}
\usepackage{xcolor}

\makeindex

\def\be#1\ee{\begin{equation}#1\end{equation}}
\def\ba#1\ea{\begin{align}#1\end{align}}
\def\bg#1\eg{\begin{gather}#1\end{gather}}

\def\shownote{1} 
\newcommand{\note}[1]{\ifthenelse{\shownote=1}{\textcolor{Red}{[[#1]]}}{}}
\def\showaddmat{1} 
\newcommand{\addmat}[1]{\ifthenelse{\showaddmat=1}{\textcolor{Gray}{[[#1]]}}{}}

\begin{document}

\title{Giant Thermal Amplification via Engineered Dissipation in a Sierpi{\'n}ski-Gasket Aharonov–Bohm Interferometer}

\author{Shubhra Shubhadarshini Mallick$^1$, Salil Bedkihal$^{2,^\dagger}$, Mattias Fitzpatrick$^2$, and Malay Bandyopadhyay$^{1,^\ast}$}
\email{malay@iitbbs.ac.in}
\email{$^\dagger$ bedkihal\_salil@yahoo.ca}
\affiliation{School of Basic Sciences, Indian Institute of Technology Bhubaneswar, Odisha, India 752050 $^1$.\\
Dartmouth Engineering Thayer School, 15 Thayer Drive, Hanover, NH 03755, USA $^2$.}



\date{ \today}

\begin{abstract}
 We propose a three-terminal thermal amplifier based on a Sierpi\'nski-gasket Aharonov--Bohm interferometer, where the third (base) terminal is realized as a floating B\"uttiker probe that acts as an engineered dissipative reservoir, exchanging energy with the conductor while carrying no net charge current. Using the nonequilibrium Green's function formalism, we demonstrate that the interplay of quantum coherence and engineered dissipation gives rise to giant, magnetic-flux-controlled thermal amplification, whereas purely coherent transport exhibits little or no amplification. We show that the amplification originates from a flux-induced cancellation of the energy-resolved thermal response of the base terminal, causing its differential heat current to vanish while finite heat currents continue to flow through the emitter and collector terminals. As a result, the thermal gain diverges without requiring resonant transmission. This interference-driven cancellation gives rise to an emergent thermal transparency, closely analogous to electromagnetically induced transparency in optical systems, where destructive quantum interference suppresses the thermal response of the base reservoir while maintaining finite heat transport through the remaining terminals. Our results establish the interplay of engineered dissipation and quantum interference as a powerful mechanism for controlling heat flow and realizing high-performance thermal amplifiers in mesoscopic conductors.
\end{abstract}


\newpage 
\maketitle
\section{Introduction}
The ability to manipulate and amplify heat flow at the nanoscale has emerged as a central challenge in quantum thermodynamics, mesoscopic physics, and energy-information technologies. Beyond its practical importance for thermal management in miniaturized electronic architectures, this problem raises fundamental questions regarding the interplay between coherence, dissipation, and energy transport in nonequilibrium quantum systems. In recent years, growing attention has been devoted to developing active thermal devices capable of controlling heat currents in a manner analogous to electronic components. Among these functionalities, thermal amplification—the ability of a small perturbation applied to a control terminal to induce a much larger change in the output heat current—has attracted particular interest because it forms the basis of thermal signal processing, thermal logic, and autonomous heat management at the nanoscale\cite{li2006negative,wang2007thermal,li2012colloquium,segal2005spin}.\\ 
\indent
Early investigations of thermal amplification were primarily carried out in classical and semiclassical systems, including nonlinear phononic lattices, anharmonic chains, and thermal rectifiers \cite{li2006negative,wang2007thermal,li2012colloquium}. These studies demonstrated that nonlinear interactions can produce controllable heat-current responses and amplification effects. However, extending these concepts to quantum-coherent electronic systems remains a significant challenge. In mesoscopic conductors, heat transport is influenced not only by dissipation and nonlinearities but also by quantum interference, phase coherence, and energy-selective transport. Quantum-interference-based transistor models have been proposed that exploit Fano–Feshbach interference and exceptional-point physics to achieve switching between low- and high-transmission states \cite{Gorbatsevich2018PT}.
These uniquely quantum effects can fundamentally modify the mechanisms through which heat currents respond to external perturbations, potentially enabling amplification mechanisms that have no classical counterpart\cite{benenti2017fundamental,whitney2014most,goldsmid2010introduction,imry1998introduction,datta1997electronic}.\\
\indent
Quantum-dot interferometers provide a versatile platform for exploring such phenomena.
Their discrete energy spectra enable energy-resolved transport, and under conditions of phase coherence, Aharonov–Bohm (AB) geometries give rise to flux-tunable quantum interference
\cite{aharonov1959significance,kobayashi2002tuning,buttiker1988coherent,bedkihal2013flux,bedkihal2013magnetic,bandyopadhyay2021flux,behera2023quantum,bedkihal2025fundamental,sridhar2026coherent}. 
Such interferometers exhibit hallmark signatures of quantum interference, including Fano resonances  \cite{sridhar2026coherent}and flux-periodic oscillations \cite{bedkihal2013flux,bedkihal2013magnetic,bandyopadhyay2021flux}, suggesting a route toward externally controllable thermal devices in which heat currents are governed by phase coherence and interference rather than solely by gradients. When such coherent transport processes are embedded within geometrically complex structures, particularly fractal lattices such as the Sierpi{\'n}ski gasket (SPG), the resulting hierarchy of interference pathways generates highly nontrivial spectral and transport properties. The self-similar nature of SPG networks provides an attractive platform for investigating how quantum interference across multiple length scales can influence thermal transport and amplification. \\
\indent
Despite substantial progress in understanding coherent heat transport in mesoscopic systems, a fundamental question remains largely unexplored: can quantum interference and fractal geometry alone generate strong thermal amplification, or is an additional dissipative mechanism required? Existing studies have largely focused on simplified scenarios, often restricted to two-terminal geometries, linear-response regimes, or models that neglect interference altogether \cite{onsager1931reciprocal,callen1998thermodynamics,benenti2011thermodynamic,saito2011thermopower}. However, nanoscale devices typically operate far from equilibrium and in environments where energy exchange, decoherence, and inelastic processes play important roles. In such situations, dissipation is not merely a source of performance degradation but can also serve as an active resource for controlling energy transport and a defining ingredient of transport.\\
\indent
A key insight is that fully coherent transport is subject to strong constraints. In the absence of inelastic processes, heat currents are determined by elastic scattering and obey Onsager reciprocity \cite{onsager1931reciprocal,callen1998thermodynamics,benenti2011thermodynamic}, which places fundamental restrictions on amplification within linear response. In contrast, inelastic scattering allows redistribution of energy between transport channels, thereby relaxing these constraints and enabling non-reciprocal functionalities such as diode- and transistor-like behaviour when time-reversal symmetry is broken. A particularly powerful and experimentally relevant framework to incorporate such effects is the voltage-probe  \cite{buttiker1986role,buttiker1986four,christen1996gauge,sanchez2004magnetic,meair2013scattering,haug2008quantum}, in which an auxiliary terminal enforces
zero net particle current while allowing energy exchange, effectively mimicking dissipation without invoking explicit many-body interactions \cite{bedkihal2013probe}. This approach provides a systematic framework for investigating how engineered dissipation modifies quantum transport and influences thermal amplification.\\
\indent
In this work, we address the following fundamental questions: (i) Can engineered dissipation serve as a resource for achieving giant thermal amplification in quantum-coherent systems? (ii) How does the fractal geometry of a Sierpi{\'n}ski gasket influence amplification characteristics? (iii) What role does magnetic-flux-controlled quantum interference play in determining amplification? (iv) What microscopic mechanism underlies the emergence of large amplification peaks in the presence of an engineered reservoir? To answer these questions, we investigate heat transport in a three-terminal Sierpi{\'n}ski-gasket Aharonov–Bohm interferometer coupled to an engineered dissipative reservoir. Using the nonequilibrium Green’s function formalism together with a self-consistent voltage-probe approach, we analyze the interplay among fractal geometry, quantum interference, and dissipation in both linear and nonlinear transport regimes.\\
\indent
\begin{figure}\label{fig:random}
    \centering
    \includegraphics[scale=0.40]{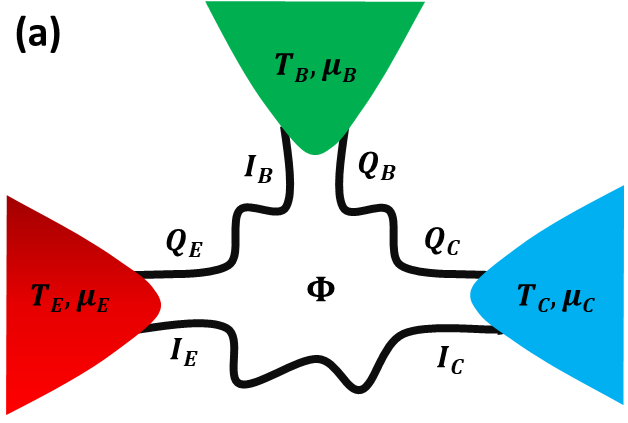}
    \includegraphics[scale=0.40]{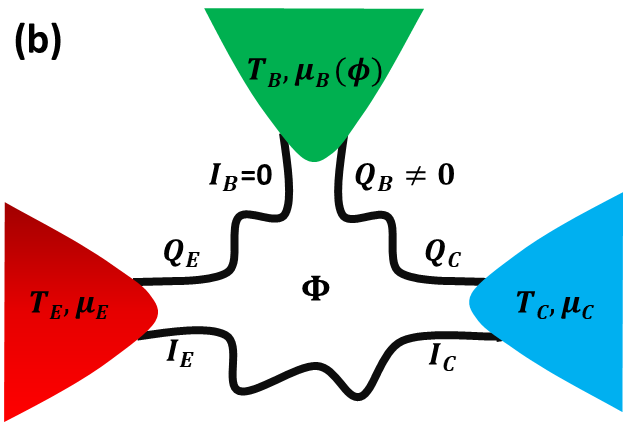}
    \caption{Schematic illustration of the three-terminal thermal transistor considered in this work. (a) All three terminals are fermionic reservoirs. (b) The emitter and collector are fermionic, while the base is modeled as a self-consistent voltage probe that allows energy exchange while enforcing $I_B=0$. A magnetic flux $\Phi$ pierces the AB ring perpendicularly.}
    \label{fig:random}
\end{figure}
We demonstrate that engineered dissipation substantially enhances thermal amplification, yielding amplification values that exceed those achievable in purely coherent systems by several orders of magnitude. Amplification can be efficiently controlled by external magnetic flux and becomes increasingly pronounced as the fractal generation of the Sierpi{\'n}ski gasket increases. We identify a microscopic mechanism responsible for this behavior: magnetic flux-controlled cancellation of the energy-resolved response of the engineered reservoir, resulting in giant amplification without requiring corresponding resonances in the transmission spectrum. This interference-driven cancellation generates an emergent thermal transparency analogous to electromagnetically induced transparency in optics, revealing a previously unexplored route toward amplification through engineered dissipation. Our results establish that the combination of fractal quantum interference and reservoir engineering provides a powerful strategy for achieving highly efficient heat-current amplification and controllable thermal signal processing in mesoscopic quantum devices.\\
\indent
The remainder of this paper is organized as follows. In Section \ref{II}, we introduce the general transport formalism and define the relevant thermal amplification measures. Section \ref{III} presents the Sierpi{\'n}ski-gasket Aharonov–Bohm model and the theoretical methodology. Section \ref{IV} discusses the amplification characteristics and the underlying interference mechanisms in the presence of engineered dissipation. Finally, Section \ref{V} summarizes our main conclusions and outlines future directions for exploiting engineered reservoirs in quantum thermal technologies. Additional technical details are provided in the appendices.

\section{General Formalism of Thermal Amplification in Dissipative Interferometers}\label{II}

In this section, we develop a general theoretical framework for thermal amplification in quantum-coherent conductors coupled to engineered reservoirs. Using the Landauer–Büttiker scattering formalism \cite{buttiker1986four,buttiker1988coherent} and nonequilibrium Green's function approach \cite{datta1997electronic}, we formulate the heat and particle transport properties of a generic three-terminal system and introduce the amplification factor together with the corresponding energy-resolved response functions. We show that engineered dissipation, implemented via a self-consistent voltage probe, enables a flux-controlled spectral-cancellation mechanism capable of producing giant thermal amplification. The framework developed here is independent of the specific device geometry and forms the basis for the analysis of the Sierpi{\'n}ski-gasket Aharonov–Bohm interferometer presented in the following section.

\subsection{Three Terminal Transport Formalism}

To establish a general framework for thermal amplification, we consider a three-terminal mesoscopic conductor connected to an emitter ($E$), a collector ($C$), and an engineered reservoir ($B$). Within the Landauer–Büttiker scattering formalism, particle and heat transport are governed by the energy-dependent transmission probabilities between the terminals \cite{Landauer1970}. This approach provides a unified description of coherent transport and engineered dissipative processes, forming the basis for analyzing thermal amplification in quantum interferometric systems.\\
\indent
The transport properties of a generic three-terminal quantum conductor can be conveniently described within the nonequilibrium Green's function (NEGF) framework. In this approach, all microscopic information about the conductor and its coupling to external reservoirs is encoded in the retarded ($G^+$) and advanced ($G^-$) Green's functions, along with the reservoir self-energies. The central quantity governing transport is the energy-dependent transmission probability between reservoirs. The transmission probability from reservoir $\nu$ to reservoir $\xi$ is given by:
\begin{equation}\label{transmission}
    T_{\nu\xi}(\omega,\phi)=Tr[\Gamma^\nu G^+(\omega,\phi)\Gamma^\xi G^-(\omega,\phi)],
\end{equation}
where $\Gamma^\nu$ and $\Gamma^\xi$ are the lead--conductor hybridization matrices, and $G^{\pm}(\omega,\phi)$ are the retarded $(+)$ and advanced $(-)$ Green's functions as derived in the appendix.
The transmission probability defined in Eq.\ref{transmission} serves as the fundamental quantity governing particle and energy transport through the conductor. In the steady state, transport arises from the exchange of carriers between reservoirs maintained at different temperatures and chemical potentials. The resulting particle current is determined by the imbalance between incoming and outgoing transmission processes, and we can express the particle currents flowing from the reservoir $\nu$ to the central system by using the Landauer–Büttiker formalism:
\begin{equation}\label{P_current}
    I_{\nu}=e \int_{-\infty}^{\infty} d\omega\sum_{\xi\ne\nu}[T_{\nu\xi}(\omega,\phi)f_{\nu}(\omega)-T_{\xi\nu}(\omega,\phi)f_{\xi}(\omega)],
\end{equation}
where,$f_{\nu(\xi)}(\omega)={[e^{(\omega-\mu_{\nu(\xi)})/T_{\nu(\xi)}}+1]}^{-1}$ is a Fermi distribution function of the reservoir $\nu(\xi)=E,C,B$ with $\mu_\nu$ and $T_\nu$ be the corresponding chemical potential and temperature, respectively. Analogously, the heat current characterizes the flow of energy measured relative to the local chemical potential of a reservoir. Within the nonequilibrium Green's function framework, the heat current extracted from the reservoir ($\nu$) is given by:
\begin{equation}\label{Q_current}
    Q_{\nu}=\int_{-\infty}^{\infty} d\omega (\omega-\mu_\nu)\sum_{\xi\ne\nu}[T_{\nu\xi}(\omega,\phi)f_{\nu}(\omega)-T_{\xi\nu}(\omega,\phi)f_{\xi}(\omega)].
\end{equation}
These expressions provide a unified description of charge and energy transport in multi-terminal quantum conductors and constitute the starting point for our analysis of thermal amplification mediated by engineered dissipation.

\subsection{Thermal Amplification and Response Functions}

We consider a three-terminal configuration comprising the emitter ($E$), collector ($C$), and engineered reservoir, which is the base ($B$) terminal. We consider two setups: one in which the reservoir (base) is engineered to act as a voltage probe and another in which the reservoir (base) is purely fermionic [Discussed in Appendix]. The central objective of a thermal amplifier is to control a large variation of an output heat current through a comparatively small perturbation applied at a control terminal. In the present three-terminal configuration, the engineered reservoir acts as the control terminal, while the emitter serves as the output channel. The efficiency of this control process is quantified by the thermal amplification factor 
\cite{PhysRevB.92.045309},
\begin{equation}\label{amplification}
\alpha=\Bigg|\frac{\partial_{T_{B}}Q_E}{\partial_{T_{B}}Q_B}\Bigg|,
\end{equation}
where $Q_E$ and $Q_B$ are the heat currents of the emitter and the base terminals, respectively, and $T_{B}$ is the base temperature. Physically, the numerator characterizes the sensitivity of the output heat current to a change in the control temperature, whereas the denominator measures the corresponding heat injected by the control terminal.\\
\indent
To elucidate the microscopic origin of amplification, it is convenient to introduce the differential thermal response functions $\chi_E = \left|\partial_{T_B}Q_E\right|,
\chi_B = \left|\partial_{T_B}Q_B\right|$, which quantify the response of the emitter and engineered reservoir heat currents to an infinitesimal variation of $T_B$. The amplification factor can then be expressed in compact form:
\begin{equation}
\alpha = \frac{\chi_{E}}{\chi_{B}}.
\end{equation}
This representation reveals that thermal amplification is fundamentally governed by the competition between two response channels. Large amplification can emerge either through an enhanced emitter response ($\chi_E$), a suppressed reservoir response ($\chi_B$), or a combination of both. In particular, giant amplification occurs when the engineered reservoir becomes nearly insensitive to thermal perturbations while the emitter retains a finite response. As we demonstrate below, such a situation can arise through interference-induced spectral cancellation, which suppresses the integrated reservoir response without significantly affecting the emitter channel.\\
\indent
The response-function formulation provides a particularly useful framework for understanding thermal amplification because it directly connects the macroscopic gain factor to the underlying energy-resolved transport processes. This perspective will serve as the basis for the amplification mechanism developed in the following subsections.

\subsection{Engineered Dissipation Through a Voltage Probe}

A central ingredient of the present work is the introduction of controlled dissipation through an engineered reservoir. In mesoscopic conductors, inelastic scattering and energy-exchange processes play a crucial role in determining transport properties, particularly under nonequilibrium conditions. Rather than introducing explicit many-body interactions, we employ the voltage-probe formalism, which provides an effective and experimentally relevant description of dissipative transport. Within this approach, the engineered reservoir acts as an auxiliary terminal that can exchange energy with the conductor while remaining electrically floating.\\
\indent
The three reservoirs are maintained at different temperatures; however, the chemical potentials of the left and right reservoirs are made distinct, i.e., $\mu_E\ne \mu_C$. Our objective is to determine $\mu_B$, which is achieved by requiring that the net particle current flowing into the engineered reservoir (base) vanishes. The defining feature of the voltage probe is the self-consistency condition that enforces the absence of net particle flow into the reservoir:
\begin{equation}\label{vp_condition}
    I_B=0.
\end{equation}

This choice enables dissipative energy-exchange processes to occur within the probe. In the linear response regime, the equation $(\ref{force})$ can be used to derive an analytical expression for $\mu_B$. In far-from-equilibrium situations, the unique chemical potential of the probe is determined numerically using the Newton-Raphson method,
\begin{equation}\label{nr_method}
    \mu_B^{(k+1)}=\mu_B^{(k)}-I_B{(\mu_B^{(k)})}\big[\frac{\partial I_B(\mu^{(k)}_B)}{\partial \mu_B}\big ]^{-1}.
\end{equation}

The current $I_B{(\mu^{(k)})}$ and its derivative are calculated from Eq. \ref{P_current} using the probe's fermi distribution with $\mu^{(k)}_B$.\\
\indent
The amplification for the fully nonlinear voltage probe case is given by,
\begin{equation}\label{amp_2}
\begin{split}
    \alpha_{\mathrm{nl}} &= \frac{|\partial_{T_B}Q_E|}{|\partial_{T_B}Q_B|}=\frac{\chi_{\mathrm{nl},E}}{\chi_{\mathrm{nl},B}},\\
    \alpha_{\mathrm{nl}} &=\frac{|\int_{-\infty}^{\infty} d\omega \chi^\mathrm{ER}_{\mathrm{nl},E}|}{|\int_{-\infty}^{\infty} d\omega \chi^\mathrm{ER}_{\mathrm{nl},B}|},
\end{split}
\end{equation}
where,
\begin{equation}
\begin{aligned}
\chi^{\mathrm{ER}}_{\mathrm{nl},E}=(\omega-\mu_E)(-T_{BE})\frac{\partial f_B(\phi)}{\partial T_B},\\
\chi^{\mathrm{ER}}_{\mathrm{nl},B}=(\omega-\mu_B(\phi))(T_{BE}+T_{BC})\frac{\partial f_B(\phi)}{\partial T_B}.
\end{aligned}
\end{equation}
Here we need to remember that in Eqs. (8) and (9), the differential response is evaluated with $\mu_B$ fixed at its self-consistent operating-point value for the specified $T_B$ and $\phi$; the voltage-probe condition determines this operating-point value through $I_B=0$. The voltage probe, therefore, serves as a minimal model of engineered dissipation, enabling the coexistence of quantum interference and irreversible energy exchange. This interplay is particularly important for thermal amplification because it relaxes the constraints imposed by purely coherent transport and provides an additional degree of control over heat currents.  In the nonlinear regime, the probe chemical potential is determined self-consistently, and the resulting dissipative dynamics give rise to the amplification mechanisms discussed in the following subsections.

\subsection{Spectral Cancellation and Emergent Thermal Transparency}

\begin{figure*}[htbp]
    \centering
    \includegraphics[scale=0.5]{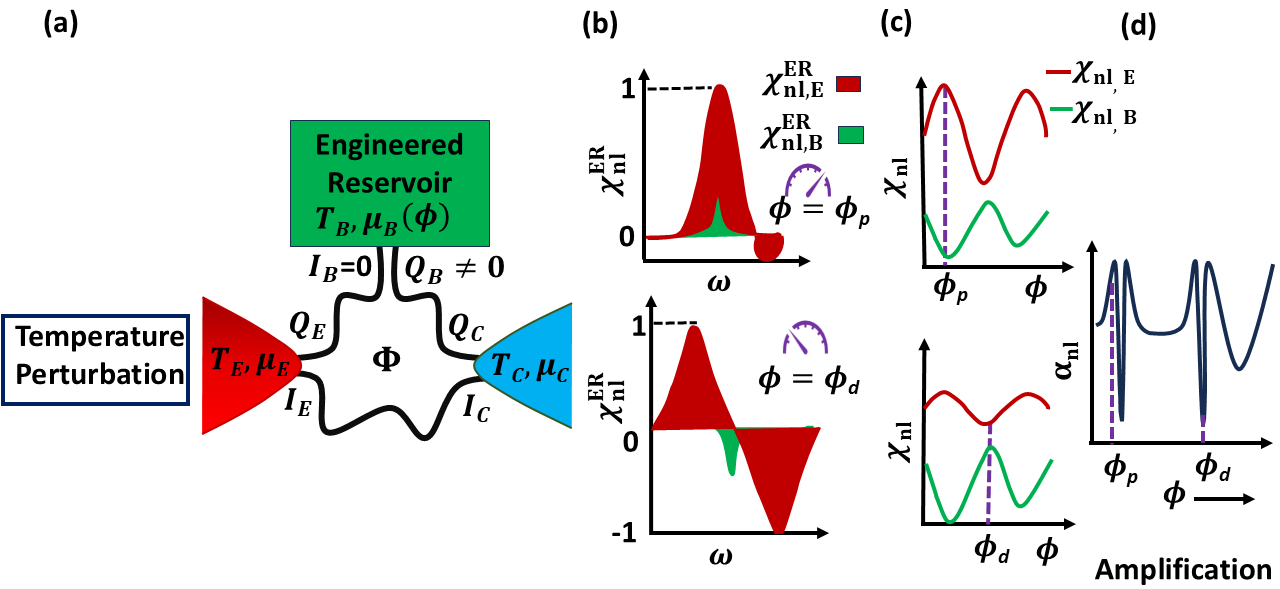}
    \caption{Schematic illustration of the mechanism underlying giant nonlinear thermal amplification. (a) A temperature perturbation is applied to the engineered reservoir (base) of a three-terminal Aharonov–Bohm interferometer operated as a voltage probe. (b) Quantum interference and engineered dissipation produce both positive and negative energy-resolved contributions to the base response, which cancel upon energy integration. (c) The emitter response remains finite while the integrated base response is strongly suppressed. (d) The resulting imbalance between the emitter and base responses produces giant thermal amplification. Here, $\phi_p$ and $\phi_d$ denote the fluxes corresponding to the amplification peak and dip, respectively.}
    \label{fig:cartoon}
\end{figure*}
A central result of the present work is that giant thermal amplification originates from the energy-resolved thermal response functions rather than from the transmission spectrum itself. In the nonlinear voltage-probe configuration, the probe is maintained in local equilibrium while satisfying the self-consistent condition $I_B=0$. Consequently, a variation of the probe temperature $T_B$ modifies not only the Fermi distribution but also the probe chemical potential $\mu_B(T_B,\phi)$, giving rise to a nonlinear thermal response that is intrinsically energy dependent.\\
\indent
Within an Aharonov–Bohm (AB) interferometer, the magnetic flux continuously redistributes the energy-dependent transmission among different interference pathways. As a result, the energy-resolved response of the engineered reservoir, $\chi^{ER}_{nl, B}(\omega)$, develops positive and negative contributions originating from different energy windows. Under suitable magnetic-flux conditions, these contributions nearly cancel after integration over energy and produces
\begin{equation}
\chi_{\mathrm{nl},B}
=
\lvert\int_{-\infty}^{\infty}
d\omega\,
\chi_{\mathrm{nl},B}^{\mathrm{ER}}(\omega)\rvert
\approx 0,
\end{equation}
while the corresponding emitter response,
\begin{equation}
\chi_{\mathrm{nl},E}
=
\lvert\int_{-\infty}^{\infty}
d\omega\,
\chi_{\mathrm{nl},E}^{\mathrm{ER}}(\omega)\rvert,
\end{equation}
remains finite. Consequently, the nonlinear amplification factor,
\begin{equation}
\alpha_{\mathrm{nl}}
=
\frac{\chi_{\mathrm{nl},E}}
{\chi_{\mathrm{nl},B}},
\end{equation}
is strongly enhanced by suppressing the integrated probe response rather than by increasing the emitter response.\\
\indent
When the conductor is realized on a Sierpi{\'n}ski-gasket (SPG) lattice, this mechanism is further reinforced by the fractal topology. The self-similar geometry generates a hierarchy of resonant states and multiple phase-coherent transport pathways spanning different length scales, producing a fragmented energy spectrum with enhanced interference. Magnetic flux, therefore, provides efficient control over the relative spectral weights of positive and negative contributions to $\chi^{ER}_{\mathrm{nl},B}(\omega)$, making the cancellation of the integrated probe response significantly more pronounced than in conventional interferometers. While leaving $\chi_{\mathrm{nl},E}^{ER}(\omega)$ comparatively unaffected, thereby providing an efficient platform for realizing giant thermal amplification through the cooperative action of quantum interference, self-consistent dissipation, and fractal spectral engineering.\\
\indent
The physical mechanism and the sequence of processes are summarised in Figure \ref{fig:cartoon}.
As illustrated in Fig. \ref{fig:cartoon}(a), a small temperature variation is applied to the engineered reservoir, which serves as the control terminal of the three-terminal interferometer. The engineered reservoir is operated as a self-consistent voltage probe satisfying ($I_B = 0$), allowing continuous energy exchange with the conductor while suppressing net particle transport. Consequently, the probe introduces controlled dissipation and also breaks transmission reciprocity. This coexistence of coherent wave propagation and dissipative energy redistribution constitutes the essential ingredient of the amplification mechanism.\\
\indent
 The microscopic origin of amplification is illustrated schematically in Fig. \ref{fig:cartoon}(b). The applied thermal perturbation generates energy-resolved response functions that contain both positive and negative spectral contributions. Quantum interference, controlled by the magnetic flux, determines the relative phases accumulated along the multiple transport pathways of the interferometer, thereby redistributing the spectral weight over different energy channels. At the same time, the voltage probe continuously exchanges energy with the conductor through its self-consistent chemical potential, modifying the occupation of these channels without introducing net charge flow. The interplay between these two effects creates conditions under which spectral contributions originating from different energy windows interfere destructively after energy integration. As a consequence, the integrated reservoir response can become strongly suppressed even though the underlying transport processes remain finite.\\
\indent
The corresponding integrated response functions are depicted schematically in Fig. \ref{fig:cartoon}(c). While the emitter response remains finite because its spectral contributions add predominantly constructively, the reservoir response approaches zero owing to the near-complete cancellation between positive and negative energy-resolved components. According to Eq. \ref{amp_2}, the amplification factor is determined by the ratio of these two integrated responses. Therefore, the simultaneous presence of a finite emitter response and a strongly suppressed reservoir response naturally produces giant thermal amplification.\\
\indent
The final consequence of this mechanism is illustrated in Fig. \ref{fig:cartoon}(d). Because the engineered reservoir responds only weakly to an external temperature perturbation while the emitter continues to exhibit a substantial heat-current variation, the output thermal signal becomes strongly amplified. This behavior may be interpreted as an emergent thermal transparency, in which the engineered reservoir becomes effectively transparent to thermal perturbations despite remaining fully coupled to the conductor and actively participating in energy transport.\\
\indent
It is important to emphasize that this thermal transparency is fundamentally different from a suppression of the transmission probability. Rather, it originates from the cancellation of the energy-resolved thermal response functions, making it a response-function phenomenon rather than a transmission phenomenon. As demonstrated in Sec. \ref{IV}, this interference-induced spectral cancellation constitutes the universal microscopic mechanism responsible for the giant thermal amplification observed in the Sierpi{\'n}ski-gasket Aharonov–Bohm interferometer.

\section{Sierpi{\'n}ski-Gasket Aharonov-Bohm Interferometer}\label{III}

\begin{figure*}[t]
    \centering
    \includegraphics[width=0.325\linewidth]{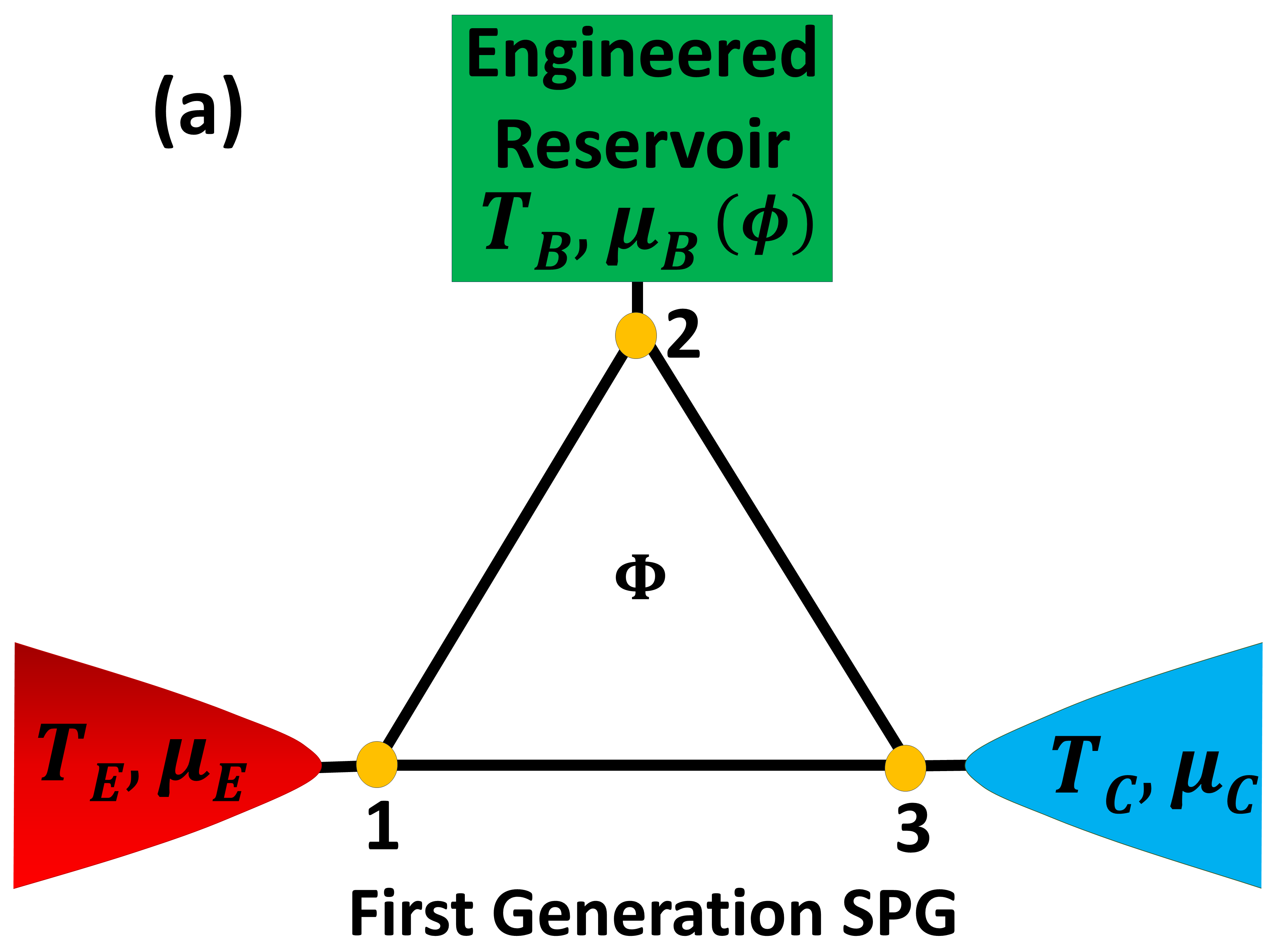}
    \includegraphics[width=0.325\linewidth]{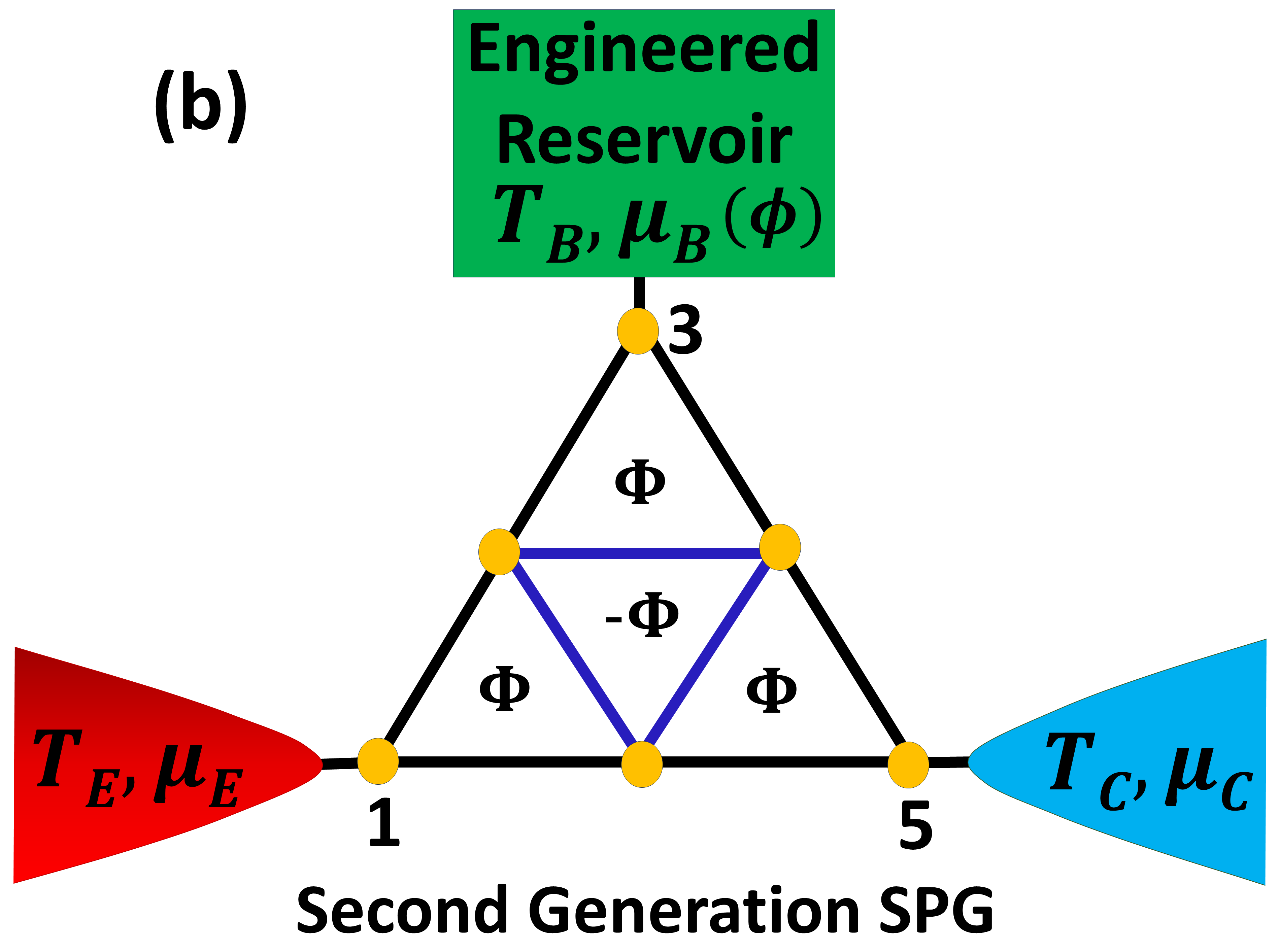}
    \includegraphics[width=0.325\linewidth]{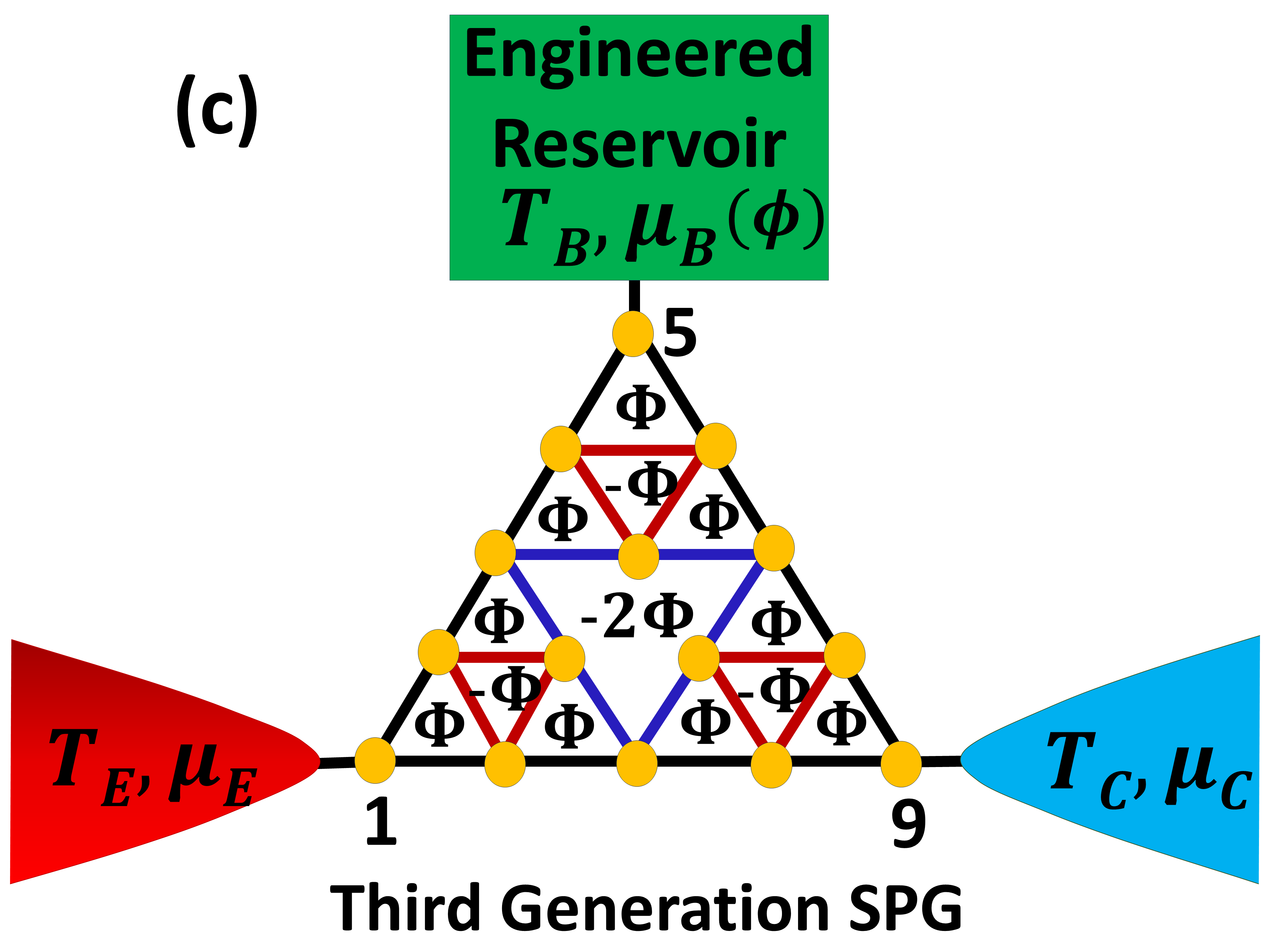}
    \caption{ Successive generations of the Sierpi{\'n}ski-gasket Aharonov–Bohm interferometer considered in this work. Quantum dots occupy the vertices of the fractal network, while the emitter ($E$), collector ($C$), and engineered reservoir ($B$) are attached to the outermost vertices. The engineered reservoir operates as a self-consistent voltage probe. The system is described by the tight-binding Hamiltonian given in Eq. \ref{SPG_system}.}
    \label{fig:3QD}
\end{figure*}
Having established the general framework for thermal amplification in dissipative interferometers, we now consider its realization in a Sierpi{\'n}ski-gasket Aharonov–Bohm (SPG-AB) interferometer. The SPG geometry provides a unique platform for exploring the interplay between fractal topology, quantum interference, and engineered dissipation owing to its self-similar structure and hierarchical transport pathways. When threaded by magnetic flux, the multiple interfering trajectories supported by the fractal network generate a rich, externally tunable energy-dependent transmission landscape.\\
\indent
Owing to the self-similar geometry of the lattice, electron waves propagate through a hierarchy of loops of different sizes, giving rise to multiple interfering paths. By successively eliminating the internal degrees of freedom of each generation, the fractal can be viewed as an effective boundary system whose energy-dependent Green function inherits the spectral features of the previous generation. At each iteration, the coupling among self-similar building blocks generates additional resonances on top of those already present, while destructive interference preserves transmission gaps. Consequently, the resonance spectrum develops recursively, with progressively finer clusters appearing on smaller energy scales. In the infinite-generation limit, this recursive refinement produces a hierarchical, Cantor-like distribution of resonances. When the lattice is connected to external reservoirs, these discrete resonances broaden into finite-width transmission peaks, so the same self-similar hierarchy becomes directly visible in the energy-dependent conductance.\\
\indent
Coupling this interferometric structure to an engineered reservoir further enables controlled energy redistribution, making the SPG-AB interferometer an ideal system for investigating the microscopic mechanisms responsible for giant thermal amplification. In the following, we introduce the model Hamiltonian and the corresponding nonequilibrium Green's function formalism used to analyze its transport properties.\\
\indent
The Sierpi{\'n}ski gasket (SPG) constitutes a paradigmatic deterministic fractal characterized by self-similarity and hierarchical connectivity. Unlike periodic lattices, the SPG possesses transport pathways distributed over multiple length scales, giving rise to rich interference phenomena and unconventional spectral properties. Such characteristics make fractal networks particularly attractive for studying coherent transport processes and their response to external perturbations.\\
\indent
The self-similar geometry recursively organizes the transmission spectrum into hierarchical resonance clusters separated by interference-induced gaps \cite{Jana2010,PhysRevB.51.9310}. Although the presence of transmission resonances is not, in general, a prerequisite for nonlinear thermal amplification, the hierarchical resonance structure substantially enhances the effect by introducing a strongly energy-selective transmission landscape. As the base temperature is varied, the thermal broadening function samples these resonance clusters, leading to a pronounced variation in the energy-weighted transmission entering $\chi_{\mathrm{nl},E}$. Meanwhile, the interference-induced transmission gaps suppress competing transport channels, limiting the corresponding variation in $\chi_{\mathrm{nl},B}$. Consequently, the emitter heat current becomes significantly more sensitive to changes in the base temperature than the heat absorbed by the base, resulting in an enhanced nonlinear thermal amplification factor $\alpha_{\mathrm{nl}}$.\\
\indent
Thus, rather than being the origin of thermal amplification, the recursive interference and the associated hierarchical resonance spectrum provide an efficient mechanism for amplifying an already existing nonlinear thermal response by increasing the sensitivity of the emitter heat current to temperature variations while maintaining relatively small heat exchange with the base.
\indent
The SPG structures considered in this work are constructed recursively by repeatedly removing the central triangle from an equilateral triangular lattice, as illustrated in Fig. \ref{fig:3QD}. The resulting generations retain the same global geometry while exhibiting increasing structural complexity. Quantum dots occupy the vertices of the fractal network and are connected through nearest-neighbor tunneling links. The emitter ($E$), collector ($C$), and engineered reservoir ($B$) are attached to the three outermost vertices of the structure.\\
\indent
 A magnetic flux $(\Phi)$ threads each elementary triangular plaquette of the network. The associated Aharonov–Bohm (AB) phase modifies the hopping amplitudes and controls the interference between different transport pathways. Consequently, the magnetic flux serves as an external tuning parameter that continuously reshapes the interference landscape of the fractal interferometer and strongly influences its transport properties \cite{PhysRevB.56.13768,PhysRevB.97.195101}.\\
\indent
For an SPG -AB interferometer, the number of sites (dots) in the $q$th generation system is given by \cite{pal2025fractal}:
\begin{equation}
    N(q)=\frac{3}{2}(1+P^q),
\end{equation}
where $P$ is the repetition factor (which is 3 for this particular fractal model) and $q=0,1,2...$ denotes the generation index. As N grows exponentially with the generation number, it provides access to increasingly complex interference networks.

We consider SPG AB interferometer \cite{bandyopadhyay2021flux,AB1, AB2,AB3,AB4,behera2023quantum,PhysRevB.85.085401}. All reservoirs are maintained at different temperatures and chemical potentials. A magnetic flux $\Phi$ pierces the setup perpendicular to it. Here, we ignore electron-electron interactions and spin degrees of freedom to construct a system that is easily solvable. Therefore, quantum dots can be modeled using a spinless electronic level, allowing us to neglect the Zeeman effect. The complete system consists of the SPG interferometer, three external reservoirs, and the corresponding tunneling couplings between them. The total Hamiltonian may therefore be written as:
\begin{equation}
    \hat{H}=\hat{H}_{SPG}+\hat{H}_{R}+\hat{H}_{SPG,R}.
\end{equation}
Here, $\hat{H}_{SPG}$ is the Hamiltonian of the subsystem for a multi-dot SPG quantum nanostructure, 
$\hat{H}_{R}$ is the reservoir Hamiltonian, and $\hat{H}_{SPG,R}$ is the Hamiltonian of the interaction between the SPG and reservoirs. The subsystem Hamiltonian corresponding to the SPG geometries within a nearest-neighbor tight-binding description is expressed as;
\begin{equation}\label{SPG_system}
   \hat{H}_{SPG}=\sum_{i=1}^N \epsilon_i \hat{d}_i^{\dagger}\hat{d}_i +\sum_{{i < j}}^N\big[ t_{ij} \hat{d}_{i}^{\dagger}\hat{d}_{j}e^{i\phi_{ij}}+H.c.\big],
\end{equation}
where $N$ denotes the number of quantum dots and $\epsilon_i$ is the energy of the $i$th dot. $\hat{d}_i^{\dagger}$ and $\hat{d}_i$ represent the creation and annihilation operators, respectively, for electrons in the corresponding quantum dots, $t_{ij}$s are tunneling strengths between the dots, and $\phi_{ij}$s are the AB phase factors. The Hamiltonians for the reservoirs, emitter ($E$), collector ($C$), and  base ($B$) are composed of non-interacting electrons and can be expressed as
\begin{equation}\label{reservoir}
    \hat{H}_{R}=\sum_{{k,\nu}\in E,C,B} \epsilon_{\nu ,k}\hat{c}_{\nu,k}^{\dagger}\hat{c}_{\nu,k}  ,
\end{equation}
where $\hat{c}_{\nu,k}^{\dagger}$ and $\hat{c}_{\nu,k}$ represent the creation and annihilation operators of the electrons in the $k^{th}$ momentum state of the reservoirs, and $\epsilon_{\nu,k}$ is the energy of $k$th state of the reservoirs. Further $\nu\in E,C,B$ represents emitter ($E$), collector ($C$), and base ($B$), respectively. The interaction Hamiltonian between the subsystem and the reservoirs is given as                       
\begin{equation}
\begin{split}
\hat{H}_{SPG,R}=\sum_{k}\bigg[V_{i,k}^{E}\hat{d}_i^\dagger\hat{c}_{E,k}+V_{i,k}^{C}\hat{d}_i^\dagger\hat{c}_{C,k}+V_{i,k}^{B}\hat{d}_i^\dagger\hat{c}_{B,k}\\
+H.c.\bigg].
\end{split}
\end{equation}
Here, $V_{i,k}^{\nu}$ represents the coupling of dots and reservoirs. 
As the magnetic flux pierces the closed loop, the electron waves gain an additional AB phase factor during their hopping between quantum dots. Magnetic flux is incorporated through the Peierls substitution in the hopping amplitudes. For each elementary triangular plaquette, the hopping between neighboring sites i and j acquires a phase $\phi_{ij}$ that obeys the following relation \cite{bandyopadhyay2021flux,behera2023quantum},
\begin{equation}
   \sum_{<ij>\in \partial\mathrm{P}} \phi_{ij}=s_{\mathrm{P}}\phi, \qquad \phi=2\pi\frac{\Phi}{\Phi_0},
\end{equation}
where $s_{\mathrm{P}}=\pm1$ specifies the orientation of the enclosed flux, $\Phi$ represents the total magnetic flux enclosed by each triangular plaquette, and $\Phi_0=h/e$ is the flux quantum. Since the dots are situated at each vertex of a triangle, we may choose the symmetric gauge as $\phi_{i,j}=s_{\mathrm{P}}\phi/3$. Here, $<ij>$ denotes a nearest-neighbor bond connecting sites $i$ and $j$, and $<ij>\in\partial\mathrm{P}$ signifies the sum over all bonds belonging to the boundary of that plaquette. For simplicity, we adopt natural units by setting $\hbar=c=k_B=1$. We convert our results to physical units in Appendix \ref{appendixD}.\\
\indent
To quantify charge and heat transport in the presence of quantum interference and engineered dissipation, we employ the nonequilibrium Green's function (NEGF) formalism. Within this framework, the influence of the reservoirs is incorporated through self-energy corrections, while all transport properties are determined by the dressed Green's functions of the SPG interferometer. The Green's functions are given as:
\begin{equation}
\begin{split}
  G^{\pm}_{SPG}(\omega)=\bigg[\omega I- H_{SPG}-\Sigma^{\pm,E}(\omega)\\-\Sigma^{\pm,C}(\omega)-\Sigma^{\pm,B}(\omega)\bigg]^{-1}.
\end{split}
\end{equation}
 Here, I is an $N \times N$ identity matrix, $\Sigma^{\pm,E}(\omega)$,  $\Sigma^{\pm,C}(\omega)$, and $\Sigma^{\pm,B}(\omega)$ are the self-energies defined in Eq.\ref{eq:A8} of Appendix \ref{appendixA}. $H_{SPG}$ is the single particle matrix corresponding to the Hamiltonian $\hat{H}_{SPG}$ in Eq. \ref{SPG_system}. In our analysis, we impose energy degeneracy for the dots $\epsilon_i=\epsilon_d$ and further consider identical symmetric interdot tunneling strength as $t_{ij}=t$ for all $i\in\lbrace1,..,15\rbrace$. Throughout this work, the reservoirs are treated within the wide-band approximation, in which the coupling matrices are energy-independent. The corresponding hybridization matrix can be defined from the relation $\Sigma^+=-i\Gamma/2$ (see Appendix \ref{appendixA}). These quantities completely determine the transmission probabilities and, consequently, the charge and heat currents discussed in Sec. \ref{II}.\\
\indent
The formalism developed above provides a complete microscopic description of transport in the SPG-AB interferometer. In the following section, we employ this framework to investigate how fractal geometry, magnetic-flux-controlled interference, and engineered dissipation cooperate to generate spectral cancellation, thermal transparency, and giant thermal amplification.
\section{Thermal Amplification in Fractal AB Interferometer: Results}\label{IV}
We now investigate how quantum interference, engineered dissipation, and fractal geometry cooperate to generate thermal amplification in the Sierpi{\'n}ski-gasket Aharonov–Bohm interferometer. The primary objective is to identify the microscopic mechanisms responsible for large amplification and to determine how they can be controlled through magnetic flux, system parameters, and fractal complexity. Particular attention is devoted to the role of the engineered reservoir, which redistributes energy dissipatively while preserving the phase coherence necessary for interference-driven transport. As demonstrated below, the interplay between these competing effects gives rise to a highly nontrivial amplification landscape characterized by giant thermal gain, strong flux tunability, and a pronounced dependence on the network's fractal generation. Most remarkably, we find that the amplification mechanism is governed not by resonant features of the transmission spectrum, but by a flux-controlled spectral cancellation of the energy-resolved reservoir response. This interference-induced cancellation generates an emergent thermal transparency, rendering the engineered reservoir nearly insensitive to thermal perturbations while maintaining a finite emitter response. The resulting imbalance between the two response channels produces exceptionally large amplification and establishes engineered dissipation as a powerful resource for controlling heat transport in quantum-coherent fractal networks. 
\begin{figure}[t!]
    \centering
    \textbf{ First Generation SPG (3QD) with Voltage Probe}\\[5pt]
    \includegraphics[width=0.49\linewidth]{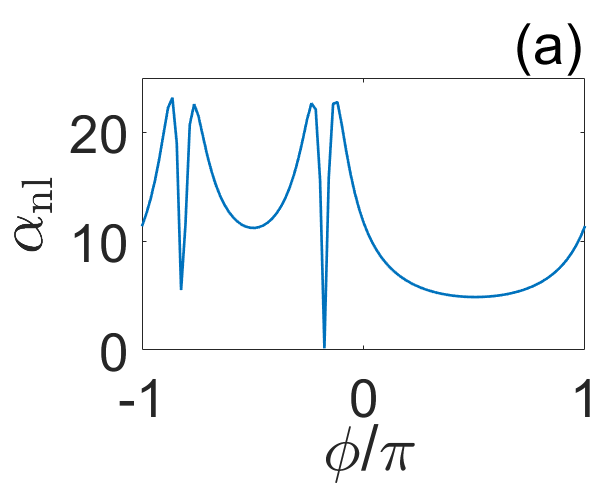}
    \includegraphics[width=0.49\linewidth]{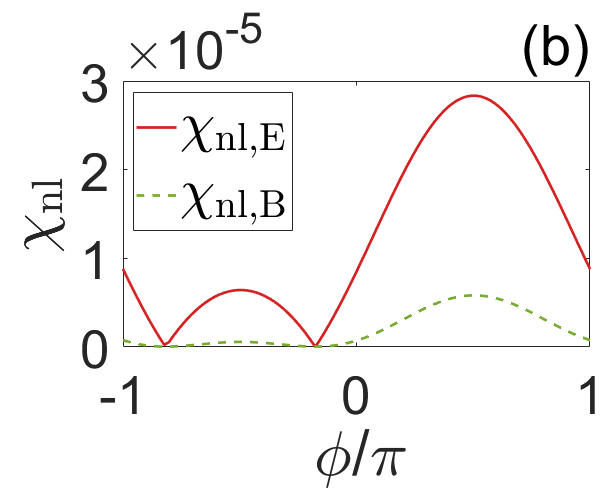}
    \includegraphics[width=0.49\linewidth]{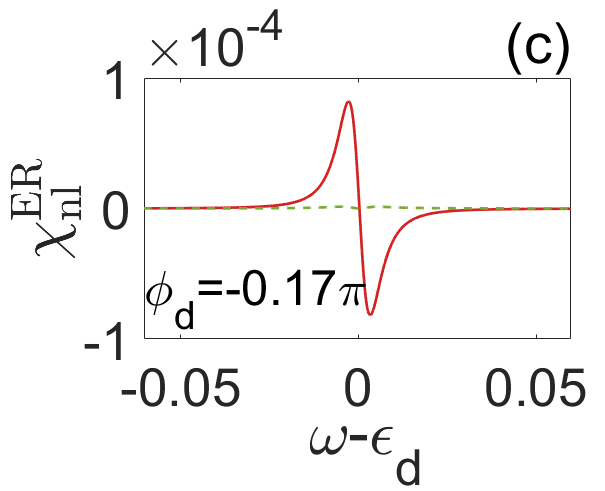}
    \includegraphics[width=0.49\linewidth]{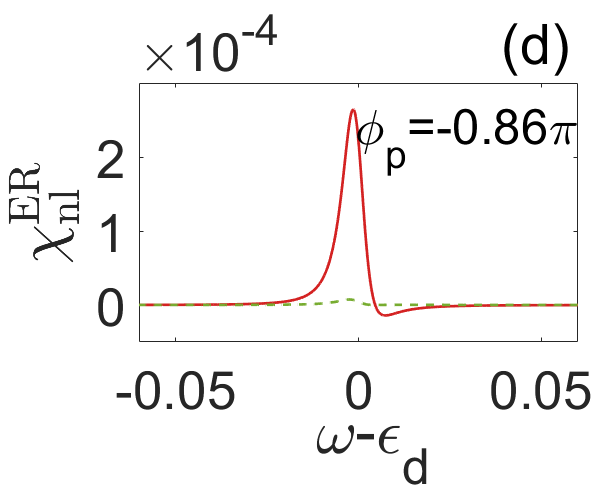}\caption{Plots of thermal amplification and thermal responses of the first-generation Sierpiński-gasket (3QD) Aharonov–Bohm thermal transistor with the base terminal operated as a voltage probe in the nonlinear regime with $t<\gamma$, and the coupling between the dots and the reservoir $\gamma_E=\gamma_C=\gamma_B=\gamma$. (a) Nonlinear thermal amplification $\alpha_{\mathrm{nl}}$ as a function of magnetic flux $\phi$ and (b) Corresponding response functions of the emitter,  $\chi_{\mathrm{nl},E}$ (red solid) and the engineered reservoir (base), $(\chi_{\mathrm{nl},B})$ (green dashed) versus magnetic flux $\phi$; Lower panel: Energy-resolved responses of emitter $(\chi^{\mathrm{ER}}_{\mathrm{nl},E})$ (red solid) and base  (engineered reservoir) $(\chi^{\mathrm{ER}}_{\mathrm{nl},B})$ (green dashed-dot) (c)  at the amplification dip flux $\phi_d$ and (d) at the amplification peak flux $\phi_p$, illustrating the flux-controlled redistribution and cancellation of the base response. Parameters used are $\gamma=0.01$ ($0.86\mu eV$), $\epsilon_d=20\gamma$ ($17.2\mu eV$), $t=0.05\gamma$ ($0.043\mu eV$), $\mu_C=-\mu_E=30\gamma$ ($25.8\mu eV$),  $T_E=60\gamma$ ($0.6K$), $T_B=30\gamma$ ($0.3K$), $T_C=10\gamma$ ($0.1K$).}
    \label{fig:QD_1V}
\end{figure}

\subsection{Flux-Controlled Thermal Amplification}

\begin{figure}[t!]
    \centering
    \includegraphics[width=1\linewidth]{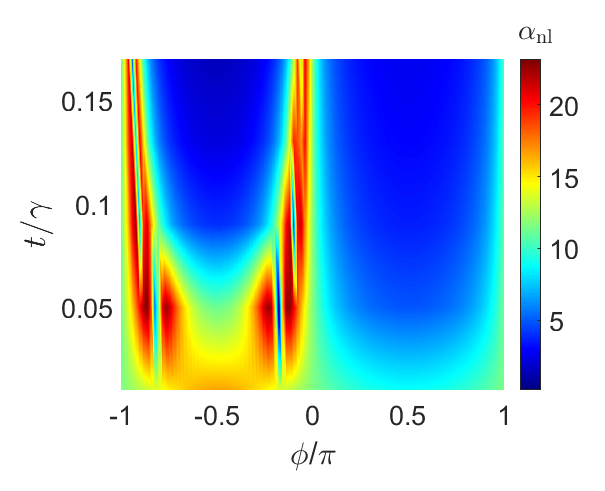}
    \caption{Amplification landscape $\alpha_{nl}$ of the first-generation SPG (3QD) thermal transistor with a voltage-probe base as a function of the interdot-to-reservoir coupling ratio $t/\gamma$ and magnetic flux $(\phi)$ in the nonlinear regime. We observe that large amplification is predominantly obtained in the weak interdot-coupling regime $t/\gamma<<1$. Parameters are the  same as in Fig.\ref{fig:QD_1V}, except for t.}
    \label{fig:heatmap_t_gamma}
\end{figure}
We begin by examining the amplification characteristics of the three-terminal SPG-Aharonov–Bohm interferometer in the nonlinear voltage-probe regime, where the engineered reservoir introduces dissipative energy redistribution while preserving quantum coherence. The choice of the system parameters is motivated by experimentally accessible quantum-dot heat engines \cite{josefsson2018quantum}. Figure \ref{fig:QD_1V} summarizes the amplification behavior of the first-generation SPG (3QD) interferometer as a function of magnetic flux. As shown in Fig. \ref{fig:QD_1V}(a), the amplification exhibits pronounced flux-dependent oscillations with several giant amplification peaks, demonstrating that the device's thermal response can be efficiently controlled via quantum interference.\\
\indent
The corresponding emitter and engineered-reservoir response functions are presented in Fig. \ref{fig:QD_1V}(b). While both responses vary periodically with the magnetic flux, the emitter response remains substantially larger than the reservoir response near the amplification maxima, thereby significantly enhancing the thermal gain. The associated energy-resolved response functions are shown in Figs. \ref{fig:QD_1V}(c) and \ref{fig:QD_1V}(d), corresponding to the minimum and maximum amplification points, and reveal that the amplification is governed by a nontrivial redistribution of spectral contributions across different energy channels. The microscopic origin of this behavior will be discussed in detail later.
\begin{figure*}
    \centering
    \includegraphics[width=0.246\linewidth]{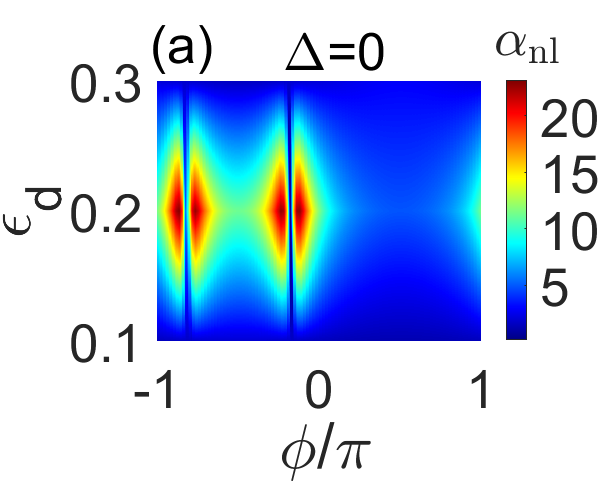}
    \includegraphics[width=0.246\linewidth]{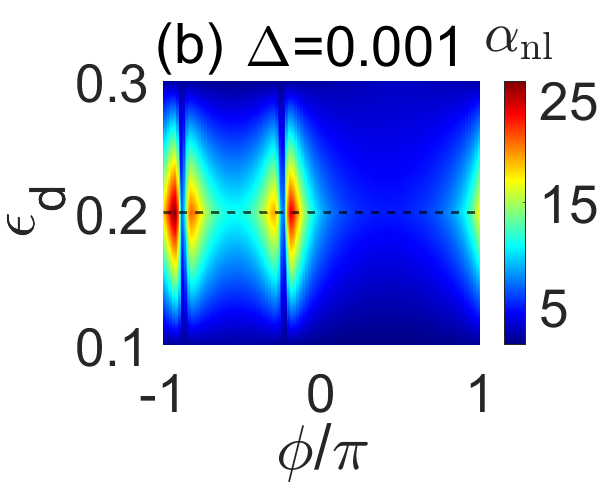}
    \includegraphics[width=0.246\linewidth]{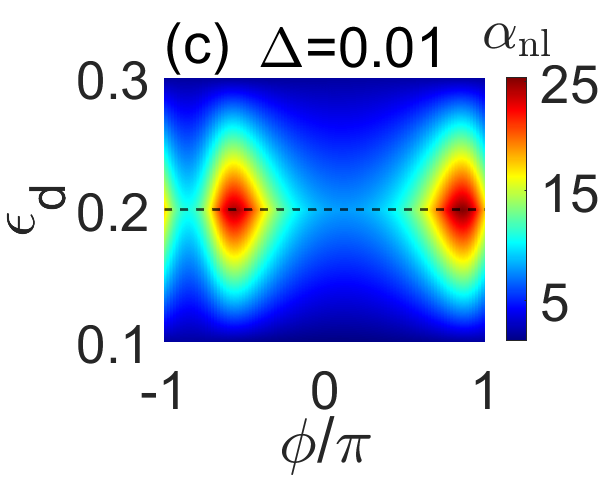}
    \includegraphics[width=0.246\linewidth]{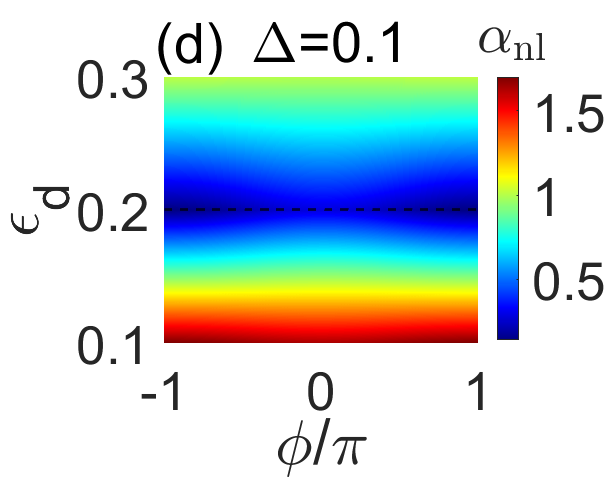}%
    \par\vspace{8pt}
   \includegraphics[width=0.246\linewidth]{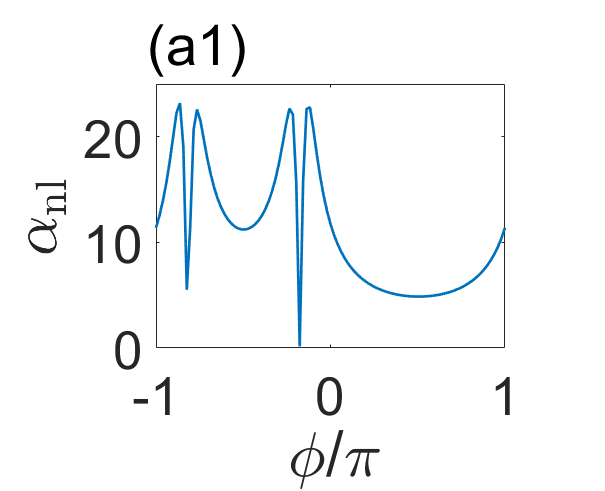}
   \includegraphics[width=0.246\linewidth]{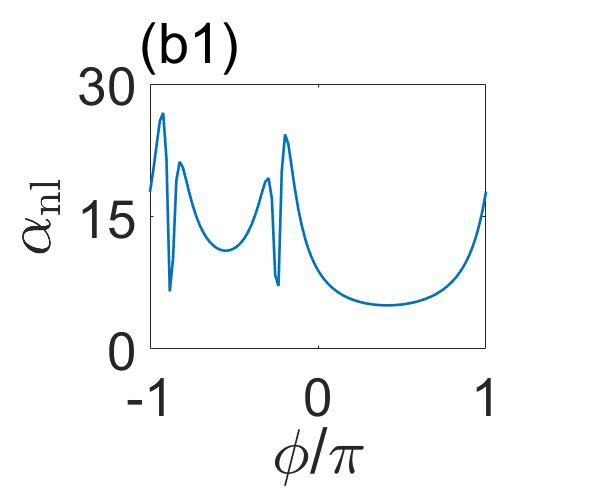}
   \includegraphics[width=0.246\linewidth]{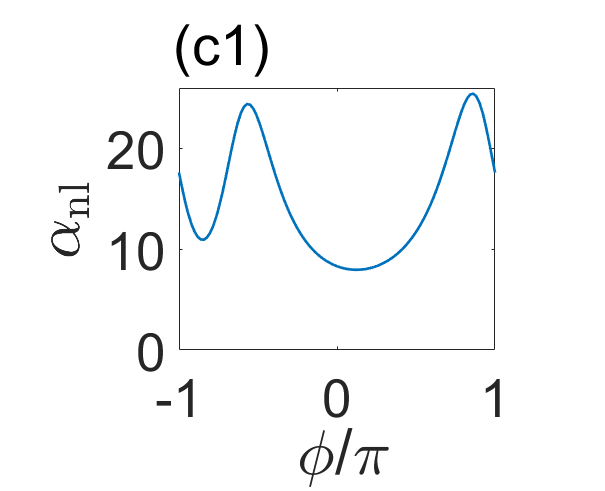}
   \includegraphics[width=0.246\linewidth]{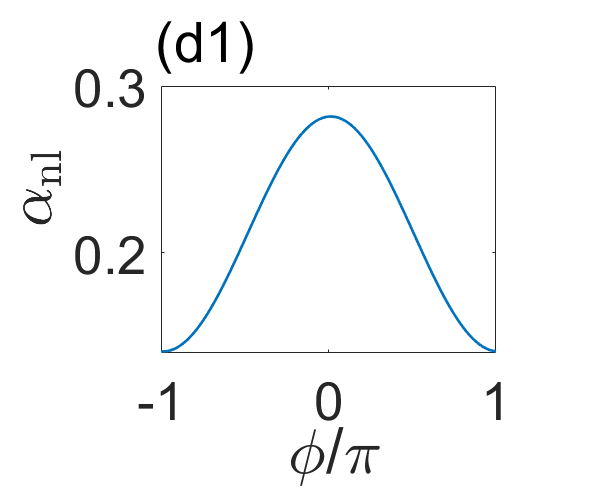}
   \caption{Dependence of the nonlinear amplification $\alpha_{nl}$ on the dot energy $\epsilon_d$ and magnetic flux $\phi$ for different level spacings $\Delta$ in the first-generation SPG (3QD) nonlinear engineered thermal transistor, where the base acts as a voltage probe. The first row (a-d) shows the corresponding amplification heatmaps for different values of $\Delta$ in the $t<\gamma$ regime. The dot energies are chosen as $\epsilon_1=\epsilon_d-\Delta$, $\epsilon_2=\epsilon_d$, $\epsilon_3=\epsilon_d+\Delta$. The second row (a1-d1) presents the corresponding line profiles of the amplification as a function of magnetic flux $\phi$, extracted from the heatmaps in the first row at $\epsilon_d=20\gamma$. Parameters are the  same as in Fig.\ref{fig:QD_1V}.}
   \label{fig:3QDa}
\end{figure*}
The robustness of the amplification is illustrated in Fig. \ref{fig:heatmap_t_gamma}, which displays the amplification landscape in the ($t/\gamma,\phi$) parameter space. Large amplification occurs predominantly in the regime ($t<\gamma$), where the interplay between quantum interference and engineered dissipation is strongest, while the amplification is considerably reduced for ($t\gtrsim\gamma$).\\
\indent
Taken together, Figs. \ref{fig:QD_1V} and \ref{fig:heatmap_t_gamma} demonstrate that the engineered reservoir exhibits large thermal amplification that can be continuously tuned by magnetic flux and optimized via system parameters. These results establish the nonlinear voltage-probe configuration as an effective platform for achieving controllable thermal amplification in fractal quantum interferometers.
\subsection{Parameter Dependence of Thermal Amplification}
\begin{figure*}[t]\label{heatmap}
    \centering
    \includegraphics[width=0.325\linewidth]{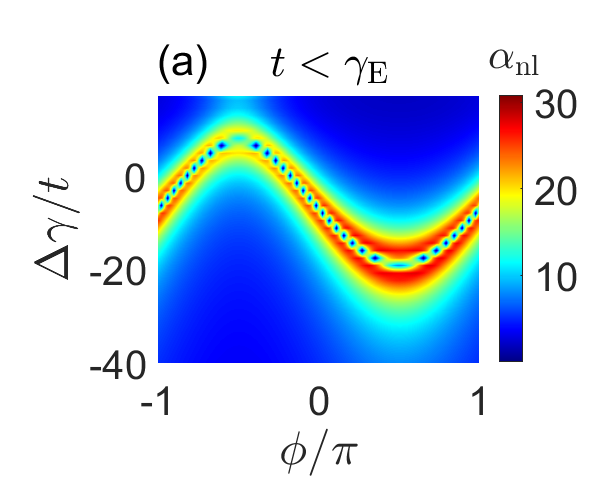}
    \includegraphics[width=0.325\linewidth]{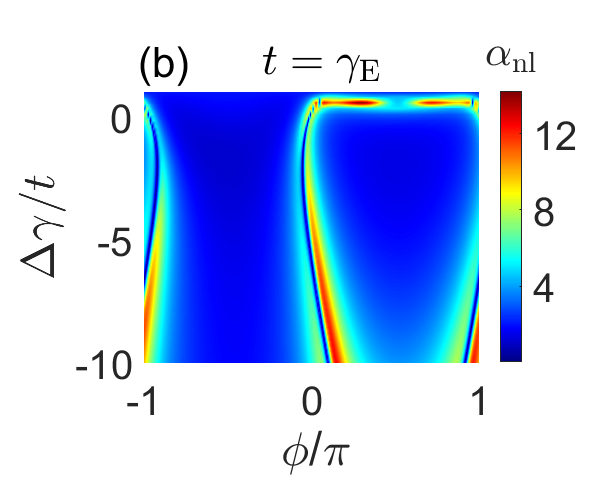}
    \includegraphics[width=0.325\linewidth]{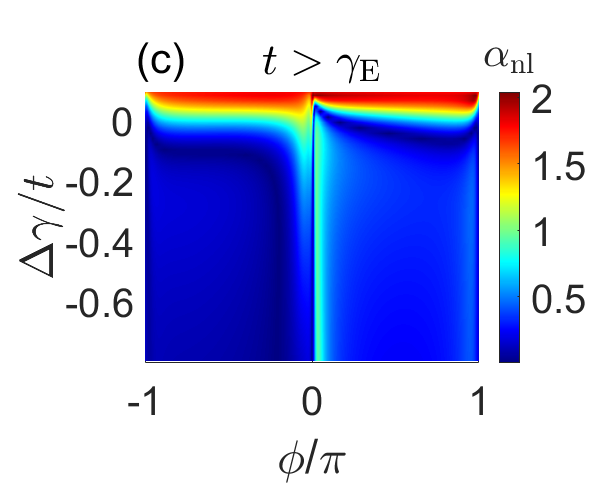}
    \caption{Amplification $\alpha_{nl}$ as a function of magnetic flux $\phi$ and the normalized emitter–collector coupling asymmetry for the first-generation SPG thermal transistor with a voltage-probe base. This illustrates the behavior of the amplification upon breaking the mirror symmetry between $\gamma_E$ and $\gamma_C$ by varying $\gamma_C$, while keeping $\gamma_E=\gamma_B=\gamma=0.01$ fixed. Other parameters are same as in Fig. \ref{fig:QD_1V}.}
    \label{fig:heatmap}
\end{figure*}
Having established the existence of giant flux-controlled thermal amplification, we now examine its dependence on the microscopic parameters of the SPG interferometer. In particular, we investigate how energy-level detuning and reservoir-coupling asymmetry influence the amplification characteristics. Such an analysis is important for assessing both the robustness of the amplification mechanism and its potential tunability in realistic device implementations.

Figure \ref{fig:3QDa} illustrates the amplification landscape as a function of the magnetic flux and the dot-level detuning ($\Delta$). The overall amplification profile remains remarkably robust to moderate detuning, with the locations of the amplification maxima shifting only slightly. At the same time, the amplification magnitude can be significantly modified by tuning the energy levels, indicating that detuning provides an additional control parameter for optimizing thermal gain. Physically, the detuning alters the relative alignment of the transport resonances, thereby modifying the interference conditions governing heat transport through the fractal network.

The influence of reservoir-coupling asymmetry is presented in Fig. \ref{fig:heatmap}. As the coupling strengths become increasingly asymmetric, the amplification landscape changes noticeably, including shifts in the positions and magnitudes of the amplification peaks. Nevertheless, giant amplification persists over a broad range of coupling configurations, demonstrating that the effect is not restricted to finely tuned symmetric structures. This robustness indicates that the amplification mechanism originates from the collective interplay between quantum interference and engineered dissipation rather than from a specific choice of device parameters.

A common feature emerging from both Figs. \ref{fig:3QDa} and \ref{fig:heatmap} is that the amplification remains highly sensitive to magnetic flux while exhibiting substantial tolerance to variations in the microscopic parameters. This behavior highlights the fundamentally interference-driven nature of the phenomenon: system parameters primarily reshape the amplification landscape, whereas the magnetic flux acts as the principal external knob controlling the thermal gain. The persistence of large amplification over an extended parameter range further suggests that the underlying mechanism is generic and can be realized under experimentally accessible conditions.

\subsection{Fractal Geometric Enhancement of Thermal Amplification}
\begin{figure}[t!]
    \centering
    \textbf{Second Generation SPG}\\[14pt]
    \includegraphics[width=0.49\linewidth]{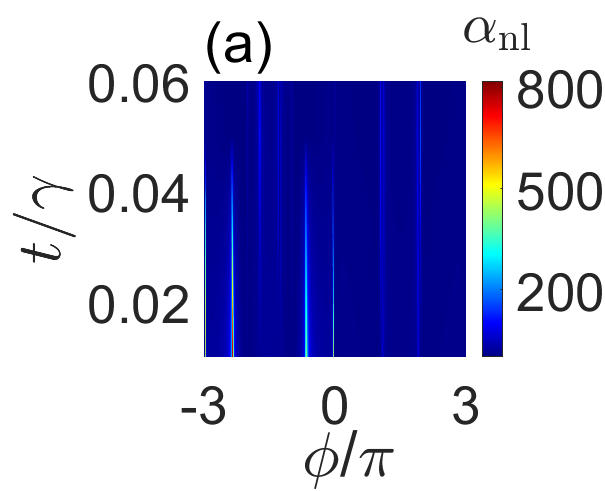}
    \includegraphics[width=0.49\linewidth]{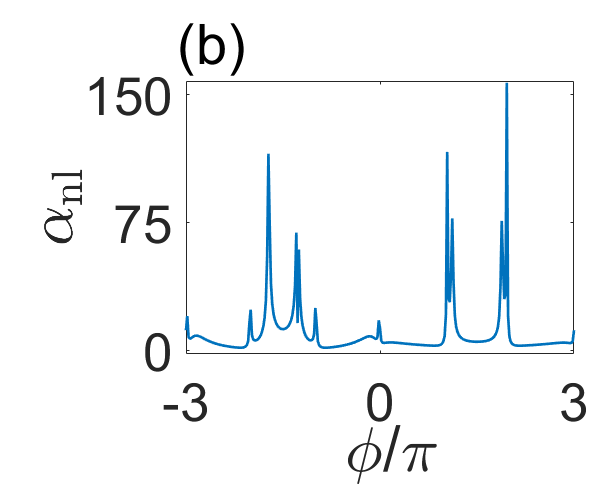}\\[14pt]
    \textbf{Third Generation SPG}\\[14pt]
    \includegraphics[width=0.49\linewidth]{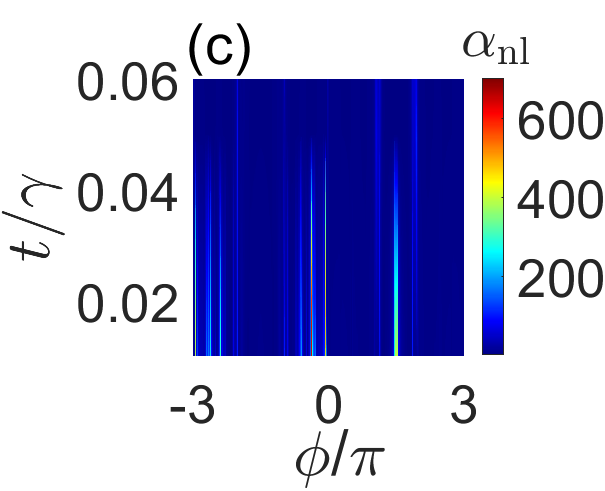}
    \includegraphics[width=0.49\linewidth]{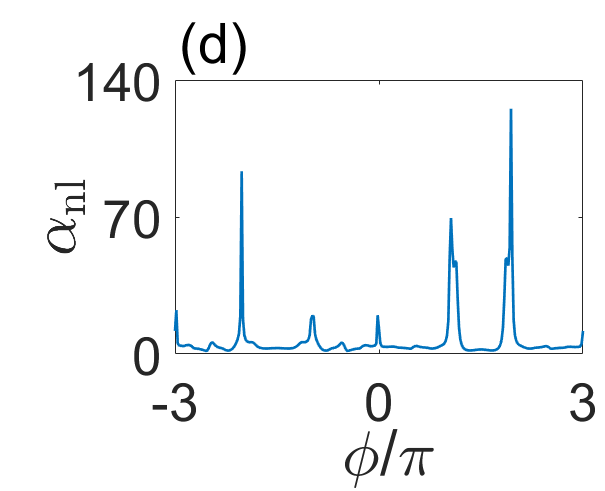}
    \caption{Fractal-generation dependence of the nonlinear thermal amplification in the SPG interferometer with a voltage-probe base. (a,c) Amplification landscapes in the $(t/\gamma,\phi)$ parameter space for the second- and third-generation SPG structures, respectively. (b,d) Corresponding amplification profiles for the second- and third-generation structures. Parameters are the same as in Fig.\ref{fig:QD_1V}.}
    \label{fig:SPG}
\end{figure}

We now investigate the influence of fractal geometry on the amplification characteristics of the SPG-Aharonov–Bohm interferometer. Unlike conventional few-site interferometers, the SPG network possesses a self-similar hierarchical structure that supports quantum-interference pathways spanning multiple length scales. This unique topology provides an ideal setting for examining how geometric complexity affects thermal amplification in the presence of engineered dissipation.\\
\indent
Figure \ref{fig:SPG} presents the amplification factor as a function of magnetic flux for different SPG generations. A clear evolution of the amplification landscape is observed as the fractal generation increases. The first-generation structure exhibits only a few well-defined amplification peaks, whereas higher generations develop a considerably richer oscillatory pattern with a larger number of amplification maxima distributed throughout the flux window. Simultaneously, the magnitude of the dominant amplification peaks increases systematically with generation, indicating that fractal complexity enhances not only the density but also the strength of the amplification response.\\
\indent
The progressive emergence of these additional amplification features reflects the hierarchical interference network inherent to the SPG geometry. Each successive generation introduces new transport loops and alternative phase-coherent pathways, thereby broadening the spectrum of interference processes. Consequently, the magnetic flux modulates an increasingly intricate interference landscape, leading to stronger fluctuations in the thermal response and more flux configurations favorable for large amplification. The increasing density of amplification peaks therefore provides direct evidence of the growing role of multiscale quantum interference in higher-generation fractal networks.\\
\indent
The results of Fig. \ref{fig:SPG} therefore demonstrate that fractal geometry is not merely a passive architectural feature but an active resource for enhancing thermal amplification. The self-similar topology amplifies quantum interference and engineered dissipation, producing increasingly complex and stronger amplification characteristics as the fractal generation increases. This generation-dependent enhancement highlights the potential of fractal quantum networks as scalable platforms for controlling heat currents and designing high-gain thermal devices.
\subsection{Giant Amplification Mechanism}
The amplification characteristics discussed above raise a fundamental question: what microscopic mechanism enables the emergence of giant thermal gain in the presence of engineered dissipation? At first sight, one might attribute the amplification peaks to resonant features of the transmission spectrum. However, a detailed analysis reveals a fundamentally different physical picture. The amplification arises from interference-induced spectral cancellation of the thermal response of the engineered reservoir, rendering it nearly insensitive to thermal perturbations while preserving a finite response in the emitter channel. As we demonstrate below, this phenomenon gives rise to an emergent thermal transparency and constitutes the central mechanism underlying the giant thermal amplification observed in the SPG-Aharonov–Bohm interferometer.\\
\indent
The physical origin of giant amplification is revealed in Fig. \ref{fig:cartoon}, which schematically illustrates the response-function framework. The energy-resolved emitter response remains predominantly of one sign and therefore contributes constructively to the integrated response. In contrast, the engineered-reservoir response exhibits substantial positive and negative spectral components distributed across different energy intervals. While each contribution is individually significant, its energy integral can be strongly suppressed by destructive interference between distinct transport channels. As a consequence, the total reservoir response approaches zero even though the underlying transport processes remain finite.\\
\indent
Direct evidence for this mechanism is provided by the energy-resolved response functions shown in Fig. \ref{fig:QD_1V}(c) and Fig.  \ref{fig:QD_1V}(d), which correspond to a minimum and a maximum of the amplification factor, respectively. At the amplification minimum, the reservoir response retains a finite integrated value because the positive and negative spectral contributions do not fully cancel. The emitter and reservoir responses, therefore, remain comparable, resulting in a relatively small amplification factor. In contrast, at the amplification maximum, the reservoir response undergoes almost complete spectral cancellation, thereby strongly suppressing its integrated contribution. Since the emitter response remains finite, the ratio of the two responses increases dramatically, producing giant thermal amplification.\\
\indent
The emergence of this cancellation can be traced to the combined action of quantum interference and engineered dissipation. The magnetic flux continuously modifies the relative phases accumulated along the numerous transport pathways supported by the SPG network, while the voltage probe redistributes energy through self-consistent dissipative dynamics. The interplay of these two ingredients generates a highly structured response landscape in which spectral contributions originating from different energy windows can interfere destructively in the reservoir channel. Importantly, this cancellation occurs at the level of the thermal response rather than the transmission function itself, explaining why the largest amplification peaks are not necessarily associated with prominent transmission resonances.\\
\indent
This phenomenon bears a conceptual resemblance to interference-induced transparency effects encountered in other areas of physics, most notably electromagnetically induced transparency in quantum optics. The analogy, however, is not rooted in the suppression of transmission. Rather, the transparency observed here arises from the cancellation of energy-resolved thermal-response contributions, which represent a fundamentally nonequilibrium transport phenomenon. The resulting amplification is therefore a manifestation of response-function engineering rather than transmission engineering.\\
\indent
The analysis above reveals a central message of the present work: engineered dissipation can act as a resource rather than a limitation for thermal transport. When combined with quantum interference and fractal topology, the voltage probe creates conditions under which the integrated reservoir response becomes nearly invisible while the output response remains finite. Giant thermal amplification thus emerges as a direct consequence of interference-driven spectral cancellation and the associated thermal transparency. This mechanism provides a general strategy for achieving high thermal gain in quantum-coherent systems and may be exploited in a broad class of mesoscopic and nanoscale thermal devices beyond the specific SPG interferometer considered here. Our analysis also reveals that amplification is enhanced in the regime where $t/\gamma << 1$. In this regime, the eigenstates are approximately delocalized between the leads, and the SPG provides multiple interference channels. The voltage probe induces a dissipative phase-breaking process, while the magnetic flux breaks time-reversal symmetry. It is interesting to note that both coherence and dissipation are simultaneously required to observe this giant thermal amplification behavior. In the limit when $t/\gamma>>1$, eigenstates are localized within SPG, and we observe the reduction in thermal amplification.
\section{Conclusions}\label{V}
In this work, we have developed a general theoretical framework for thermal amplification in quantum-coherent multiterminal conductors with engineered dissipation and demonstrated its realization in a Sierpi{\'n}ski-gasket Aharonov–Bohm interferometer. Employing the nonequilibrium Green's function formalism together with a self-consistent voltage-probe approach, we showed that introducing an engineered dissipative reservoir dramatically enhances thermal amplification beyond that achievable in purely coherent systems. The amplification is highly tunable by magnetic flux, remains robust against variations in the microscopic system parameters, and is further strengthened by the hierarchical interference network of the fractal geometry.\\
\indent
More importantly, our analysis uncovers a microscopic mechanism fundamentally different from conventional resonance-based transport phenomena. We demonstrate that giant thermal amplification originates from interference-induced spectral cancellation of the energy-resolved response of the engineered reservoir. The cooperative action of quantum interference and dissipative energy redistribution suppresses the integrated reservoir response while preserving a finite emitter response, thereby producing exceptionally large thermal gain. This mechanism gives rise to an emergent thermal transparency, in which the engineered reservoir becomes effectively insensitive to thermal perturbations despite remaining fully coupled to the conductor. The amplification, therefore, arises from the engineering of thermal response functions rather than from the transmission spectrum itself, establishing a new paradigm for controlling heat transport in coherent quantum systems.\\
\indent
Our results further reveal that the self-similar topology of the Sierpi{\'n}ski gasket provides an intrinsic geometric advantage for thermal amplification. As the fractal generation increases, the hierarchy of phase-coherent transport pathways enriches the interference landscape and substantially enhances the amplification characteristics. This demonstrates that fractal geometry is not merely a structural feature but an active resource that cooperates with engineered dissipation to achieve high thermal gain.\\
\indent
The proposed mechanism is experimentally accessible with present-day mesoscopic technologies. Quantum-dot Aharonov–Bohm interferometers defined in semiconductor two-dimensional electron gases or gate-defined quantum-dot arrays provide a natural platform for realizing the proposed geometry. A floating metallic contact operated under the Büttiker voltage-probe condition can emulate the engineered reservoir, while magnetic flux can be controlled through an external perpendicular magnetic field. Local electronic temperatures may be established using integrated nanoscale heaters, and heat currents can be inferred from sensitive thermoelectric or noise-based thermometry techniques. Recent advances in semiconductor quantum-dot circuits, nanofabricated interferometers, and multiterminal thermal transport measurements make the proposed amplification mechanism experimentally feasible.\\
\indent
Beyond the specific SPG architecture considered here, the physical principle established in this work is expected to be applicable to a broad class of coherent mesoscopic systems, including quantum-dot networks, molecular junctions, topological interferometers, and other fractal or hierarchical quantum lattices. More generally, our results demonstrate that engineered dissipation can be transformed from an unavoidable source of decoherence into a functional resource for controlling energy transport. We anticipate that the concepts of spectral cancellation and emergent thermal transparency introduced here will stimulate new approaches to thermal signal processing, quantum caloritronics, and the design of high-performance quantum thermal devices.
\section{Acknowledgments}
S.S.M. acknowledges financial support received from IIT Bhubaneswar in the form of an Institute Research Fellowship.
\appendix
\section{Appendix A: Equations of Motion}\label{appendixA}
The model setup of the Sierpi{\'n}ski-gasket Aharonov–Bohm (SPG-AB) interferometer is described in detail in section \ref{III}. We now solve this model and compute the observables in the non-equilibrium steady state. Since the model is non-interacting, we can apply the NEGF approach to evaluate its steady-state properties \cite{pooja,wang2014nonequilibrium}. The NEGF technique has been widely used in recent years to study transport properties in mesoscopic systems and molecular junctions \cite{fransson2010non}. We use the equation-of-motion (EOM) approach in the derivations \cite{dharsen}. This approach involves solving the Heisenberg equation of motion for the reservoir variables, then inserting the resulting expressions back into the EOM for the subsystem (dots) variables. We obtain a subsystem quantum Langevin equation as follows:
\begin{equation}\label{eq:A1}
\begin{aligned}
     \frac{d\hat{d}_i}{dt}=-i\Bigg[\epsilon_i \hat{d}_i+\sum_{j\ne i} t_{ij} \hat{d}_j e^{i\phi_{ij}}\Bigg] -i\sum_{\nu=E,C,B} \hat\eta^{\nu}_{i}(t)\\
    -i \sum_{j,\nu=E,C,B}\int_{t_0}^t \Sigma_{i,j}^{\nu,+} (t-t')\hat{d}_{j}(t')dt' .
\end{aligned}
\end{equation}
Here, the index $i$ is used to identify the dots, while $\hat{\eta}^E_i$,   $\hat{\eta}^C_i$ and $\hat{\eta}^B_i$ denote the noise contribution on the subsystem by the emitter, collector, and base (engineered reservoir), respectively.\\
The expressions are
\begin{equation}
    \begin{aligned}
        \hat{\eta}^{E}_{i}=i\sum_{k} V_{i,k}^{E}  g_{Ek}^{+}(t-t_0)\hat{c}_{Ek}(t_0),\\
        \hat{\eta}^{C}_{i}=i\sum_{k} V_{i,k}^{C}  g_{Ck}^{+}(t-t_0)\hat{c}_{Ck}(t_0),\\
        \hat{\eta}^{B}_{i}=i\sum_{k} V_{i,k}^{B}  g_{Bk}^{+}(t-t_0)\hat{c}_{Bk}(t_0).
    \end{aligned}
\end{equation}
For the isolated reservoirs, the retarded Green's functions are given as follows:
\begin{equation}\label{eq:A3}
    \begin{aligned}
        g_{Ek}^{+}(t)=-i e^{-i\epsilon_{Ek}t}\theta(t),\\
        g_{Ck}^{+}(t)=-i e^{-i\epsilon_{Ck}t}\theta(t),\\
        g_{Bk}^{+}(t)=-i e^{-i\epsilon_{Bk}t}\theta(t).
    \end{aligned}
\end{equation}
For the initial condition, the total density matrix is to be factorized as $\rho_T(t_0)=\rho_E\otimes\rho_B\otimes\rho_C\otimes\rho(t_0)$, with empty dots and reservoirs prepared in a grand canonical state
\begin{equation}
    \hat{\rho}_{\nu}=\frac{e^{-(\hat{H}_\nu-\mu_\nu \hat{N})/T_\nu}}{Tr[e^{-(\hat{H}_\nu-\mu_\nu \hat{N})/T_\nu}]},
\end{equation}
Where $T_\nu$ and $\mu_\nu$ are the temperatures and chemical potentials of the reservoirs with $\nu=E,C,B$. The reduced density matrix $\rho$ represents the state of the subsystem. Using the initial conditions, we can obtain the noise correlation as follows:
\begin{equation}
    \begin{aligned}
        \Big\langle \hat{\eta}_i^{\dagger E}(t)  \hat{\eta}_{i'}^{\dagger E}(\tau)\Big\rangle=\sum_{k} V_{i,k}^{E^*} e^{i\omega_k(t-\tau)} V_{i',k}^{E} f_E(\omega_k),\\
        \Big\langle \hat{\eta}_i^{\dagger C}(t)  \hat{\eta}_{i'}^{\dagger C}(\tau)\Big\rangle=\sum_{k}  V_{i,k}^{C^*}  e^{i\omega_k(t-\tau)}  V_{i',k}^{C} f_C(\omega_k),\\
        \Big\langle \hat{\eta}_i^{\dagger B}(t)  \hat{\eta}_{i'}^{\dagger B}(\tau)\Big\rangle=\sum_{k}  V_{i,k}^{B^*}  e^{i\omega_k(t-\tau)}  V_{i',k}^{B} f_B(\omega_k),
    \end{aligned}
\end{equation}
with the Fermi function $f_\nu=[e^{(\omega-\mu_\nu)/T_\nu}+1]^{-1}$ for the reservoir $\nu=E,C,B$ and $\mu_\nu$ and $T_\nu$ are the corresponding chemical potential and temperature, respectively. According to the Heisenberg picture, the expectation value of an observable A can be obtained as $\langle \hat{(A)}\rangle=Tr_T[\rho_T(t_0)\hat{A(t)}]$, taking the trace over all degrees of freedom. The properties of the steady state are obtained by taking the limits $t_0\to-\infty$ and $t_0\to \infty$. Now we can obtain the Fourier transform of Eq.\eqref{eq:A1} using the convolution theorem with the convention $\overset{\sim}{d}_i=\int_{-\infty}^{\infty} dt d_i(t) e^{i\omega t}$ and $\overset{\sim}{\eta}_i^\nu(\omega)=\int_{-\infty}^{\infty}dt \eta_i^\nu(t)e^{i\omega t}$ and this will be shown in matrix form as
\begin{equation}
    \overset{\sim}{d}_i(\omega)=\sum_{j} G^+_{i,j}(\omega)[\overset{\sim}{\eta}_{j}^E(\omega)+\overset{\sim}{\eta}_{j}^C(\omega)+\overset{\sim}{\eta}_{j}^B(\omega)].
\end{equation}
Here, the retarded Green's function is given as
\begin{equation}
\begin{aligned}
G^{+}_{SPG}(\omega)
= \Big[ & \omega I - H_{SPG} - \Sigma^{+,E}(\omega) \\
        & - \Sigma^{+,C}(\omega)- \Sigma^{+,B}(\omega) 
\Big]^{-1},
\end{aligned}
\end{equation}
where $I$ is the identity matrix of size $(N\times N)$, and $N$ denotes the number of QDs in the AB ring.  The advanced Green's function is given by the conjugate transpose of the retarded Green's function, $G^-(\omega)={[G^+(\omega)]}^\dagger$. The self-energies are given as
\begin{equation}\label{eq:A8}
    \begin{aligned}
        \Sigma^{\pm,E}(\omega)=\sum_k V_{i,k}^E g_{Ek}^\pm(\omega)V^{E^*}_{i,k},\\
        \Sigma^{\pm,C}(\omega)=\sum_k V_{i,k}^C g_{Ck}^\pm(\omega)V^{C^*}_{i,k},\\
        \Sigma^{\pm,B}(\omega)=\sum_k V_{i,k}^B g_{Bk}^\pm(\omega)V^{B^*}_{i,k}.
     \end{aligned}
\end{equation}

Here, $g_{Ek}^\pm(\omega)$, $g_{Ck}^\pm(\omega)$ and $g_{Bk}^\pm(\omega)$  are the Fourier transforms of Eq.\eqref{eq:A3}. In the wide-band limit (WBL), where the density of states (DOS) of metallic lead (reservoir) is energy independent, the real component of the self-energy vanishes. The hybridization matrix can be written from the relation $\Sigma^+=-i\Gamma/2$
\begin{equation}
\begin{aligned}
    \Gamma^E_{i,i'}=2\pi \sum_{k} V_{i',k}^{E^*} V_{i,k}^{E}\delta(\omega-\omega_k),\\
    \Gamma^C_{i,i'}=2\pi \sum_{k} V_{i',k}^{C^*} V_{i,k}^{C}\delta(\omega-\omega_k),\\
    \Gamma^B_{i,i'}=2\pi \sum_{k} V_{i',k}^{B^*} V_{i,k}^{B}\delta(\omega-\omega_k).
  \end{aligned}  
\end{equation}
$V_{i,k}^{E}$, $V_{i,k}^{C}$ and $V_{i,k}^{B}$ may be taken as real constants, independent of both the level index and reservoir state, resulting in $\Gamma^E_{i,i'}=\gamma_E$, $\Gamma^C_{i,i'}=\gamma_C$ and $\Gamma^B_{i,i'}=\gamma_B$, where $\gamma_\nu$ (energy independent) represents the coupling between the dots and the metallic leads (reservoir). For simplicity, we assume degenerate dot energies $\epsilon_i=\epsilon_d$ and set $t_{ij}=t$, from which the retarded Green's function is derived.

\section{Appendix B : Onsager Coefficients}\label{appendixB}
For a non-interacting system, the particle current ($I_\nu$) and the heat current ($Q_\nu$) flowing from reservoir $\nu$ to $\xi$ (where $\nu$,$\xi$=$E,C,B$) can be expressed using the Landauer-Buttiker formalism as:
\begin{equation}
    I_{\nu}=e\int_{-\infty}^{\infty} d\omega\sum_{\xi\ne\nu}[T_{\nu\xi}(\omega,\phi)f_{\nu}(\omega)-T_{\xi\nu}(\omega,\phi)f_{\xi}(\omega)],
\end{equation}
\begin{equation}
    Q_{\nu}=\int_{-\infty}^{\infty} d\omega (\omega-\mu_\nu)\sum_{\xi\ne\nu}[T_{\nu\xi}(\omega,\phi)f_{\nu}(\omega)-T_{\xi\nu}(\omega,\phi)f_{\xi}(\omega)],
\end{equation}
where $f_{\nu(\xi)}(\omega)={[e^{(\omega-\mu_{\nu(\xi)})/T_{\nu(\xi)}}+1]}^{-1}$ is a Fermi distribution function of the reservoir $\nu(\xi)=E,C,B$ with $\mu_\nu$ and $T_\nu$ be the corresponding chemical potential and temperature, respectively and $T_{\nu\xi}$ is the transmission probability from $\nu$ to $\xi$ terminal. The Onsager coefficients are obtained by performing a linear expansion of the particle current ($I_\nu$) and the heat current ($Q_\nu$).
\subsection{Linear Fermionic Thermal Transistor} 
A three-terminal thermal transistor that operates with broken time-reversal symmetry is illustrated in Fig.\ref{fig:3QD}(a). A scattering region exposed to an external magnetic flux $\Phi$ is connected to three distinct fermionic reservoirs labeled as emitter ($E$), collector ($C$), and base ($B$)  that can exchange both heat and particles with the system. We take the collector reservoir as the reference by setting $(\mu_C,T_C)=(\mu,T)$, and express the chemical potential and temperature of reservoir $\nu=E,C,B$ as $\mu_\nu=\mu+\Delta\mu_\nu$,$T_\nu=T+\Delta T_\nu$, respectively. If $|\Delta\mu_\nu/T|<<1$ and $|\Delta T_\nu|/T<<1$, then our system operates in the linear response regime. Under the above condition, the particle and heat currents are described in Eqs. (\ref{P_current}) and (\ref{Q_current}) can be expressed by linear expansion and obtain the linear Onsager relation between the thermodynamic fluxes (particle and heat currents) and forces (chemical potential and temperature biases) as $\bm{J=MA}$:  
\begin{equation}\label{onsagar_1}
\begin{pmatrix}
    I_E\\
    Q_E\\
    I_B\\
    Q_B
\end{pmatrix}=
{\begin{pmatrix}
      M_{11} & M_{12} & M_{13} & M_{14} \\
      M_{21} & M_{22} & M_{23} & M_{24}\\
      M_{31} & M_{32} & M_{33} & M_{34}\\
      M_{41} & M_{42} & M_{43} & M_{44}
\end{pmatrix}}
\begin{pmatrix}
    A_1\\
    A_2\\
    A_3\\
    A_4
\end{pmatrix},
\end{equation}
where $A_1=(\mu_E-\mu_C)/(eT_C)$, $A_2=(T_E-T_C)/T_C^2$, $A_3=(\mu_B-\mu_C)/(eT_C)$, and  $A_4=(T_B-T_C)/T_C^2$ are the generalized forces. Here, M is a $4\times 4$ Onsager matrix and its elements $M_{ij}$ are called the Onsager coefficients, and we will suppress $\phi$ in the Onsager coefficients to keep the notation simple, unless necessary. The detailed forms of the Onsager coefficients are provided below.\\
\indent
From Eq.(\ref{amplification}) amplification factor for the fermionic thermal transistor in the linear response regime is given as follows:
\begin{equation}\label{linear_fermionic}
    \alpha_{\mathrm{l}'}=\Bigg|\frac{M_{24}}{M_{44}}\Bigg|=\frac{\chi_{{\mathrm{l}'},E}}{\chi_{{\mathrm{l}'},B}},
\end{equation}
where index  ${\mathrm{l}'}$ denotes the linear response regime when all the reservoirs are fermionic.
The above equation only depends on two Onsager coefficients $M_{24}$ and $M_{44}$.

\subsection{Linear Voltage Probe Thermal Transistor }

The probe parameters ($T_B$,$\mu_B$) are tuned to prevent particle flow. The voltage probe (base) induces dissipative inelastic scattering processes in the system by enforcing a zero net particle current into the probe $(I_B=0)$, while permitting a nonzero heat current $(Q_B \ne 0)$ from the probe (base). Thus, from Eq. (\ref{onsagar_1}), we can obtain $A_3$ as follows:
\begin{equation}\label{force}
    A_3=-\frac{(M_{31}A_1+M_{32}A_2+M_{34}A_4)}{M_{33}}.
\end{equation}
Substituting Eq. (\ref{force}) in Eq. (\ref{onsagar_1}), and after some calculation, we can express the currents in terms of a $(3\times 3)$ reduced Onsager matrix $M'$ as follows \cite{PhysRevB.87.205420,PhysRevB.85.085401}:
\begin{equation}
\begin{pmatrix}
    I_E\\
    Q_E\\
    Q_B
\end{pmatrix}=
{\begin{pmatrix}
      M_{11}' & M_{12}' & M_{13}' \\
      M_{21}' & M_{22}' & M_{23}'\\
      M_{31}' & M_{32}' & M_{33}'
\end{pmatrix}}
\begin{pmatrix}
    A_1\\
    A_2\\
    A_4
\end{pmatrix},
\end{equation}
where the reduced Onsager coefficients $M_{ij}'$ are expressed as follows:
\begin{equation}
\begin{aligned}
M_{11}'=\frac{M_{33}M_{11}-M_{13}M_{31}}{M_{33}},
M_{12}'=\frac{M_{33}M_{12}-M_{13}M_{32}}{M_{33}},\\
M_{13}'=\frac{M_{33}M_{14}-M_{13}M_{34}}{M_{33}},
M_{21}'=\frac{M_{33}M_{21}-M_{23}M_{31}}{M_{33}},\\
M_{22}'=\frac{M_{33}M_{22}-M_{23}M_{32}}{M_{33}},
M_{23}'=\frac{M_{33}M_{24}-M_{23}M_{34}}{M_{33}},\\
M_{31}'=\frac{M_{33}M_{41}-M_{43}M_{31}}{M_{33}},
M_{32}'=\frac{M_{33}M_{42}-M_{43}M_{32}}{M_{33}},\\
M_{33}'=\frac{M_{33}M_{44}-M_{43}M_{34}}{M_{33}}.
\end{aligned}
\end{equation}
Amplification for the voltage probe in the linear response regime is given as, 
\begin{equation}
        \alpha_\mathrm{l}=\Bigg|\frac{M_{23}'}{M_{33}'}\Bigg|=\Bigg|\frac{M_{33}M_{24}-M_{23}M_{34}}{M_{33}M_{44}-M_{43}M_{34}}\Bigg|=\frac{\chi_{\mathrm{l},E}}{\chi_{\mathrm{l},B}}.
\end{equation}
\indent
The Onsager coefficients are given as follows:
\begin{equation}
    \begin{aligned}
       M_{11} &=e^2T\int_{-\infty}^{\infty}d\omega(-d_\omega f) (T_{EC}+T_{EB}),\\
       M_{12} &=eT\int_{-\infty}^{\infty}d\omega(-d_\omega f)(\omega-\mu)(T_{EC}+T_{EB})=M_{21},\\
       M_{13} &=e^2T\int_{-\infty}^{\infty}d\omega(-d_\omega f) (-T_{BE}),\\
       M_{14}&=eT\int_{-\infty}^{\infty}d\omega(-d_\omega f)(\omega-\mu) (-T_{BE})=M_{23},\\
       M_{22}&=T\int_{-\infty}^{\infty}d\omega(-d_\omega f){(\omega-\mu)}^2 (T_{EC}+T_{EB}),\\
        M_{24}&=T\int_{-\infty}^{\infty}d\omega(-d_\omega f){(\omega-\mu)}^2 (-T_{BE}),\\
        M_{31}&=e^2T\int_{-\infty}^{\infty}d\omega(-d_\omega f) (-T_{EB}),\\
        M_{32}&=eT\int_{-\infty}^{\infty}d\omega(-d_\omega f){(\omega-\mu)} (-T_{EB})=M_{41},\\
        M_{33}&=e^2T\int_{-\infty}^{\infty}d\omega(-d_\omega f) (T_{BE}+T_{BC}),\\
        M_{34}&=eT\int_{-\infty}^{\infty}d\omega(-d_\omega f) {(\omega-\mu)} (T_{BE}+T_{BC})=M_{43},\\
        M_{42}&=T\int_{-\infty}^{\infty}d\omega(-d_\omega f){(\omega-\mu)}^2 (-T_{EB}),\\
         M_{44}&=T\int_{-\infty}^{\infty}d\omega(-d_\omega f) {(\omega-\mu)}^2 (T_{BE}+T_{BC}).
        \end{aligned}
\end{equation}
Here $(d_\omega f=-[4T\cosh^2\big(\frac{\omega-\mu}{2T}\big)]^{-1})$ represents the first-order derivative of the Fermi distribution function with energy.

\subsection{Fully Nonlinear Fermionic Thermal Transistor}
We can find the amplification for the fully nonlinear fermionic transistor by using the heat current $(Q_E)$ flowing from the reservoir emitter to the central system and the heat current $(Q_B)$ flowing from the reservoir base to the central system.
\begin{equation}\label{amp_1}
\begin{split}
    \alpha_{\mathrm{nl}'} &=\frac{|\partial_{T_{B}}Q_E|}{|\partial_{T_{B}}Q_{B}|}=\frac{\chi_{{\mathrm{nl}'},E}}{\chi_{{\mathrm{nl}'},B}}\\
   \alpha_{\mathrm{nl}'} &=\frac{|\int_{-\infty}^{\infty} d\omega \chi^{\mathrm{ER}}_{{\mathrm{nl}'},E}|}{|\int_{-\infty}^{\infty} d\omega \chi^{\mathrm{ER}}_{{\mathrm{nl}'},B}|}
\end{split}
\end{equation}
where,
\begin{equation}
\begin{aligned}
\chi^{\mathrm{ER}}_{{\mathrm{nl}'},E}=(\omega-\mu_E)(-T_{BE})\frac{\partial f_B}{\partial T_B},\\
\chi^{\mathrm{ER}}_{{\mathrm{nl}'},B}=(\omega-\mu_B)(T_{BE}+T_{BC})\frac{\partial f_B}{\partial T_B},\\
\frac{\partial f_B}{\partial T_B}=\bigg(\frac{\omega-\mu_B}{T_B^2}\bigg) \frac{\exp(\frac{\omega-\mu_B}{T_B})}{(\exp(\frac{\omega-\mu_B}{T_B})+1)^2},
\end{aligned}
\end{equation}
where ${\mathrm{nl}'}$ denotes the nonlinear response regime when all the reservoirs are fermionic. The amplification in Eq.(\ref{amp_1}) depends on two transmission functions $T_{BE}$ and $T_{BC}$. The transmission functions $T_{BE}$ and $T_{BC}$ are different for different geometries.
\section{Appendix C : Physical Observables}\label{appendixC}
\subsection{ First Generation SPG (3QD) AB Thermal Transistor}
By applying the quantum Langevin equation approach as discussed in Refs.\cite{dharsen}, the retarded Green function for our system can be derived as follows(for details, please see Appendix \ref{appendixA})
\begin{equation}
\renewcommand{\arraystretch}{1.5}
G^+_{3QD}(\omega)=
{\begin{pmatrix}
    \omega-\epsilon_d+i\frac{\gamma_E}{2}  &  -t e^{i\phi/3}  &  -t e^{-i\phi/3}\\
  -t e^{-i\phi/3}  &  \omega-\epsilon_d+i\frac{\gamma_B}{2}  &  -t e^{i\phi/3}\\
  -t e^{i\phi/3}  &  -t e^{-i\phi/3}  &   \omega-\epsilon_d+i\frac{\gamma_C}{2}
\end{pmatrix}}^{-1}.
\end{equation}
In matrix form $\Gamma^E$, $\Gamma^C$, and $\Gamma^B$ are given by\\
$$\Gamma^E_{3QD}={\begin{pmatrix}
    \gamma_E & 0 & 0\\
    0 & 0 & 0\\
    0 & 0 & 0
\end{pmatrix}},
\Gamma^C_{3QD}={\begin{pmatrix}
    0 & 0 & 0\\
    0 & 0 & 0\\
    0 & 0 & \gamma_C
\end{pmatrix}},$$
$$\Gamma^B_{3QD}={\begin{pmatrix}
    0 & 0 & 0\\
    0 & \gamma_B & 0\\
    0 & 0 & 0
\end{pmatrix}}.$$
\\
\indent
The advanced Green's function is defined as the conjugate transpose of matrix $G_{3QD}^+$;
\begin{equation}
    G_{3QD}^-(\omega)={[G_{3QD}^+(\omega)]}^\dagger.
\end{equation}
\begin{figure}[t!]
    \centering
    \textbf{ First Generation SPG (3QD) Without Voltage Probe}\\[5pt]
    \includegraphics[width=0.49\linewidth]{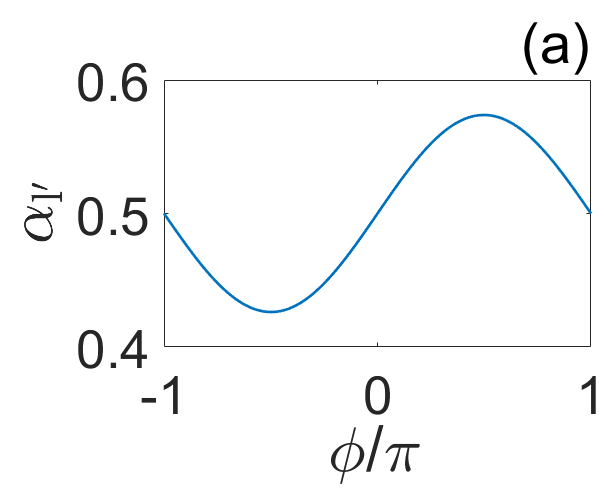}
    \includegraphics[width=0.49\linewidth]{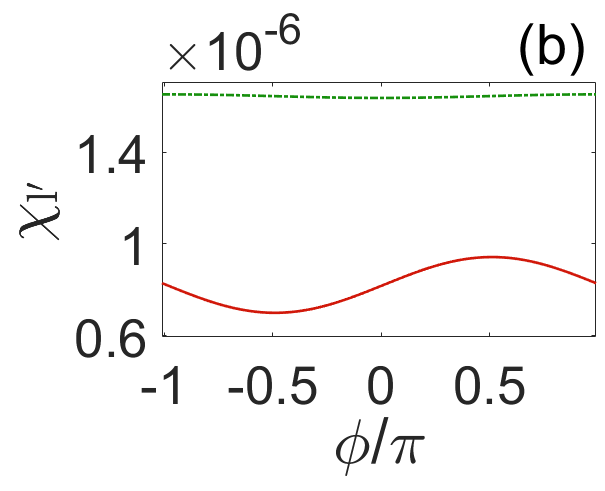}
    \includegraphics[width=0.49\linewidth]{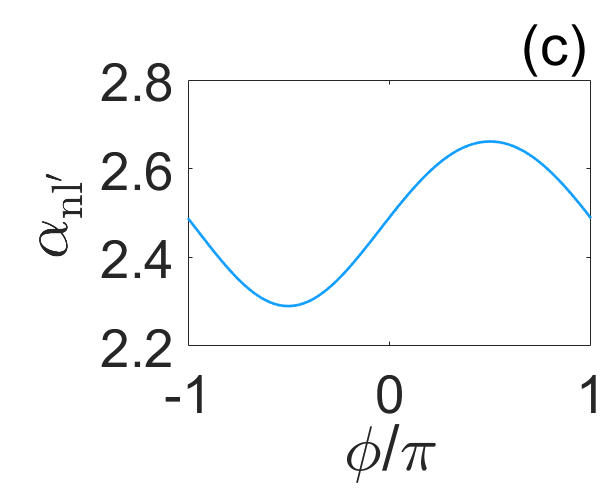}
    \includegraphics[width=0.49\linewidth]{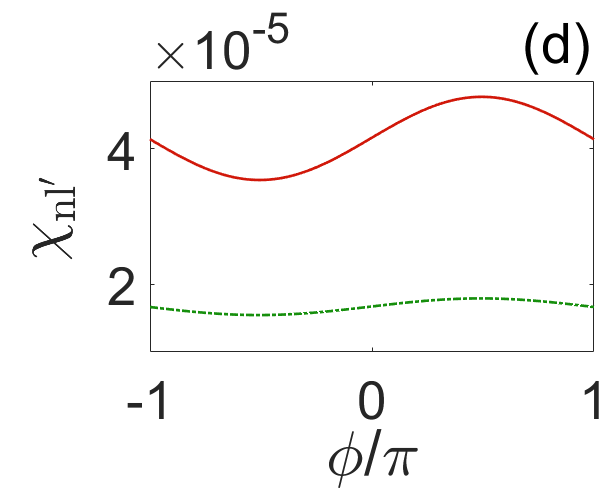}

    \includegraphics[width=0.49\linewidth]{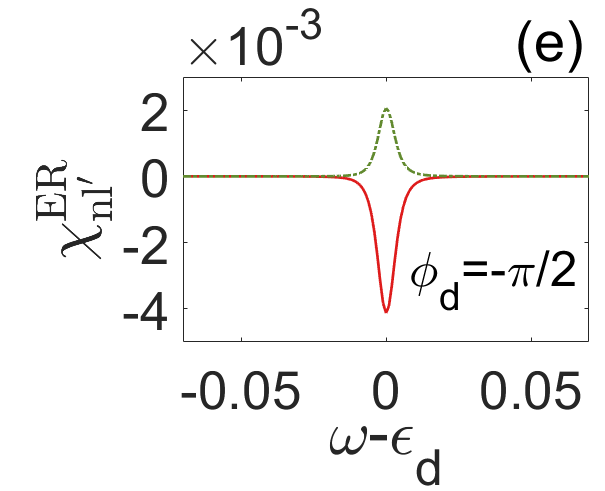}
    \includegraphics[width=0.49\linewidth]{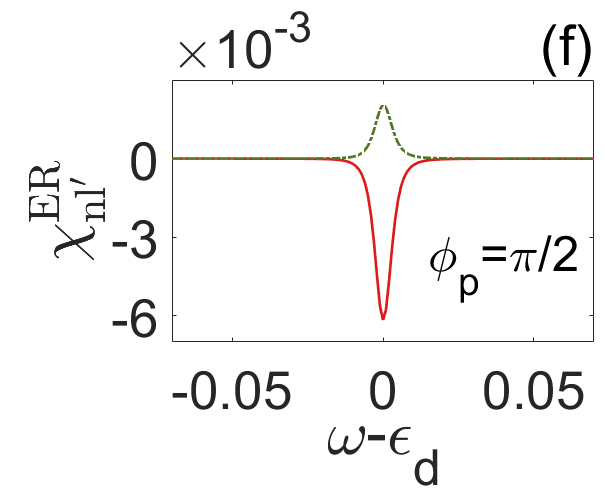}
    \caption{Thermal amplification in the first-generation SPG (3QD) interferometer with a purely fermionic base. Solid red and dashed-dotted green curves denote the emitter and base responses, respectively. (a,b) Linear-response amplification and corresponding emitter/base responses as functions of magnetic flux. (c,d) Nonlinear amplification and corresponding responses. (e,f) Energy-resolved emitter and base responses at the fluxes corresponding to the amplification dip $\phi_d$ and the peak $\phi_p$, respectively. The linear response results in the first panel (a,b) corresponding to $\mu_E=30\gamma (25.8\mu eV)$, $\mu_C=30.4\gamma (26.14\mu eV)$, $\mu_B=30.2\gamma$ ($25.97\mu eV$), $T_E=50\gamma$ ($0.5K$),  $T_C=45\gamma$ ($0.45K$), $T_B=46.5\gamma$ ($0.465K$), with all remaining parameters as in Fig. \ref{fig:QD_1V}. The second and third panels use the same parameters as Fig. \ref{fig:QD_1V} and $\mu_B=10\gamma (8.6 \mu eV)$.}
    \label{fig:QD_1F}
\end{figure}

The transmission functions are given as
\begin{equation}
  T_{3QD}^{BE}=\frac{
  \begin{aligned}
  & \gamma^2\bigg[t^4+t^2({(\omega-\epsilon_d)}^2+\frac{\gamma^2}{4}) \\ & + t^3(2(\omega-\epsilon_d)\cos\phi 
  +\gamma\sin\phi)\bigg]
\end{aligned}
}{\Delta},
\end{equation}
\begin{equation}
T_{3QD}^{BC}=\frac{ 
\begin{aligned}
& \gamma^2\bigg[t^4+t^2({(\omega-\epsilon_d)}^2+\frac{\gamma^2}{4})
\\
& +t^3(2(\omega-\epsilon_d)\cos\phi-\gamma\sin\phi)\bigg]
\end{aligned}
}{\Delta},
\end{equation}
\begin{equation}
T_{3QD}^{BE}+T_{3QD}^{BC}=2\frac{
\begin{aligned}
    &\gamma^2\bigg[t^4+t^2({(\omega-\epsilon_d)}^2+\frac{\gamma^2}{4})\\
    & +2t^3(\omega-\epsilon_d)\cos\phi\bigg]
    \end{aligned}}{\Delta},
\end{equation}
\begin{equation}
\begin{aligned}
    \Delta={\bigg[(\omega-\epsilon_d)({(\omega-\epsilon_d)}^2-3(t^2+\frac{\gamma^2}{4}))-2t^3\cos\phi\bigg]}^2\\
    +\frac{\gamma^2}{4}{\bigg[3{(\omega-\epsilon_d)}^2-(\frac{\gamma^2}{4}+3t^2)\bigg]}^2.
\end{aligned}
\end{equation}

Transmission functions $T_{BE}$ and $T_{BC}$  only contribute to amplification for first-generation SPG (3QD) configurations.
\subsection{Fermionic Base: Absence of Amplification in Linear Response and Emergence of Weak Amplification in Nonlinear Response}

\begin{figure}
    \centering
    \includegraphics[width=0.49\linewidth]{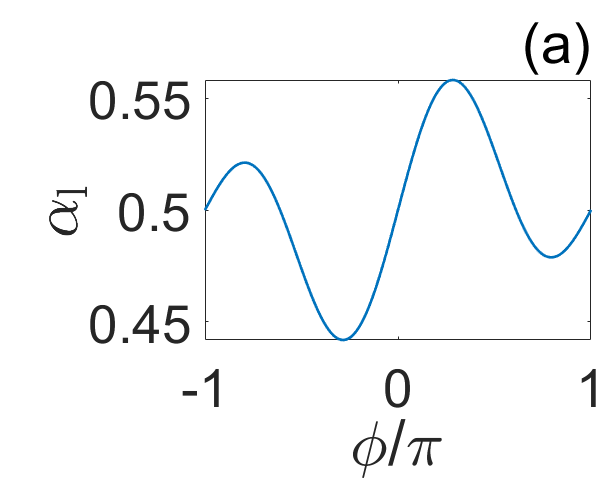}
    \includegraphics[width=0.49\linewidth]{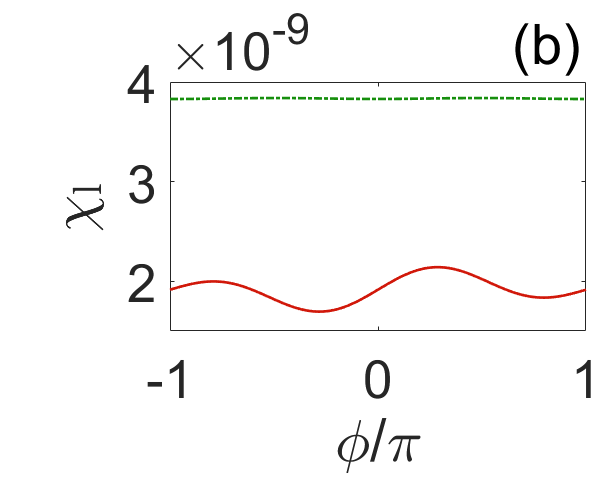}
    \caption{Plots show the first-generation SPG (3QD) thermal transistor (linear response condition) in $(t < \gamma)$ regime. Here, the base acts as an engineered voltage probe. (a) shows the amplification $\alpha_{\mathrm{l}}$ as a function of $\phi$ and (b) shows the response function $\chi_{\mathrm{l}}$ as a function of $\phi$. The solid red line represents the emitter response function$(\chi_{\mathrm{l},E})$ and the dashed-dot green line corresponds to the base response function $(\chi_{\mathrm{l},B})$. Parameters used are  $\mu_E=30\gamma (25.8\mu eV)$, $\mu_C=30.4\gamma (26.14\mu eV)$, $T_E=50\gamma$ ($0.5K$), $T_C=45\gamma$ ($0.45K$), $T_B=46.5\gamma$ ($0.465K$), with all remaining parameters as in Fig. \ref{fig:QD_1V}.}
    \label{fig:QD_1LV}
\end{figure}
\begin{figure*}
    \centering
    \includegraphics[width=0.32\linewidth]{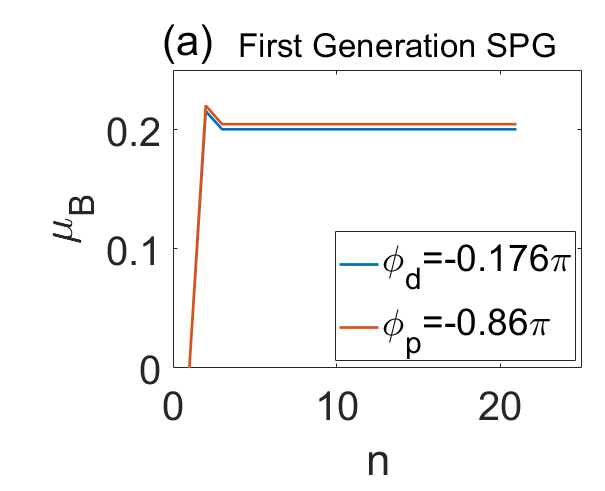}
    \includegraphics[width=0.32\linewidth]{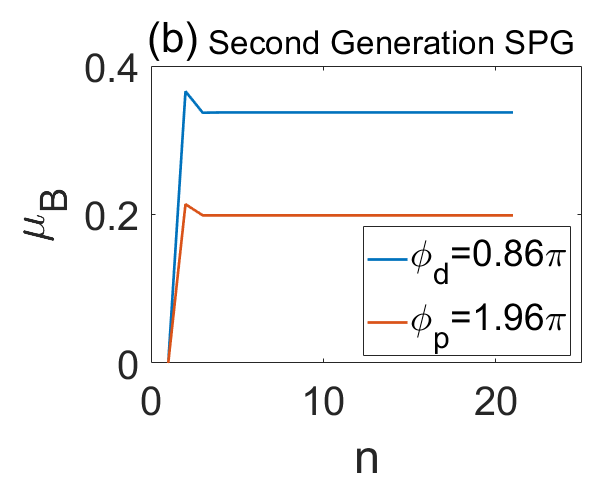}
    \includegraphics[width=0.32\linewidth]{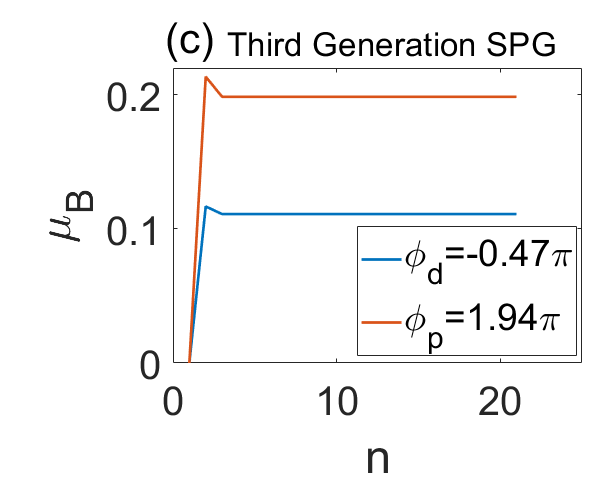}
    \includegraphics[width=0.32\linewidth]{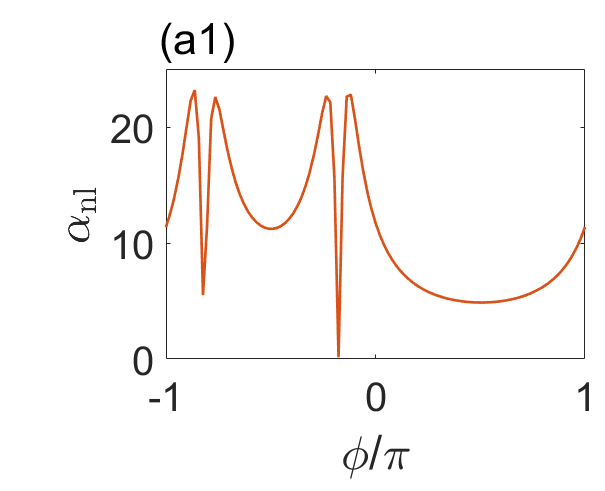}
    \includegraphics[width=0.32\linewidth]{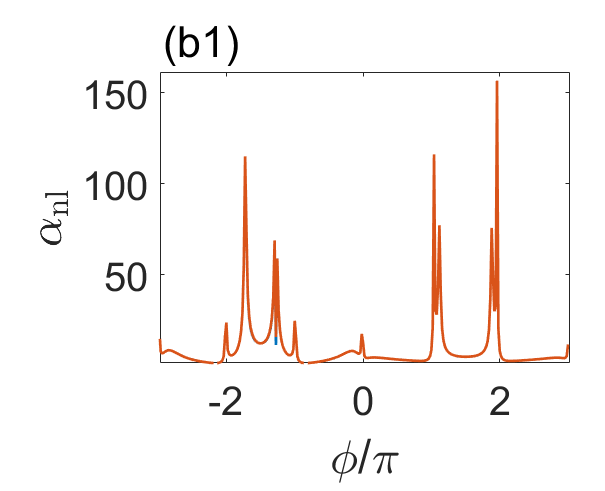}
    \includegraphics[width=0.32\linewidth]{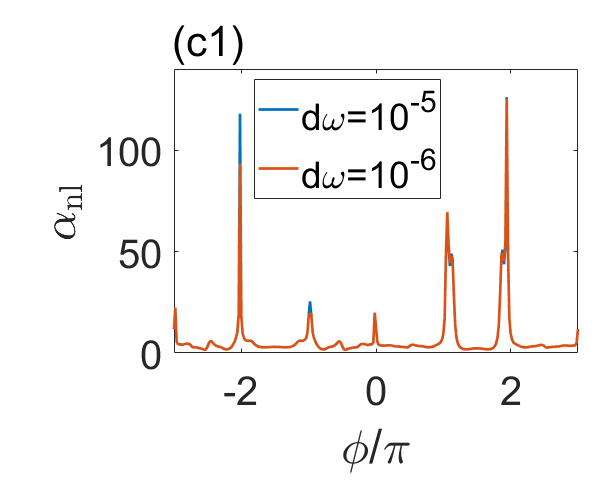}
    \caption{Convergence of the self-consistent voltage-probe calculation for the first-, second-, and third-generation SPG structures. (a–c) Convergence of the probe chemical potential $\mu_B$ as a function of the iteration number $n$ at the fluxes corresponding to the amplification dip $\phi_d$ (blue solid) and the amplification peak $\phi_p$ (red solid). Convergence analysis of the self-consistent probe method. (a1–c1) Convergence of the calculated amplification $\alpha_{nl}$ as the magnetic-flux resolution/window is refined. The self-consistent chemical potential is obtained by enforcing $I_B=0$. The parameters used are the same as those in Fig. \ref{fig:QD_1V}.}
    \label{fig:convergence}
\end{figure*}

We first consider the first-generation SPG (3QD) interferometer configuration with purely fermionic equilibrium reservoirs. In the linear-response regime Eq. [\ref{linear_fermionic}], the amplification factor remains below unity for all magnetic flux values in the strong-coupling regime ($t<\gamma$), as shown in Fig. [\ref{fig:QD_1F}(a)], indicating the absence of transistor action. The emitter and base responses [Fig. \ref{fig:QD_1F}(b)] remain comparable throughout the flux range, reflecting the constraints imposed by Onsager reciprocity and purely elastic transport. Consequently, no significant asymmetry develops between the emitter and base channels, preventing amplification.\\
\indent
In the nonlinear regime [Eq. \ref{amp_1}], finite amplification emerges for all flux values [Fig. \ref{fig:QD_1F}(c)]. As shown in Fig. \ref{fig:QD_1F}(d), the emitter response exceeds the base response, yielding $\alpha>1$ and demonstrating transistor behavior. This enhancement originates from nonlinear corrections to the Fermi-Dirac distributions.\\
\indent
The microscopic origin of this behavior is illustrated by the energy-resolved responses in Figs. \ref{fig:QD_1F}(e) and \ref{fig:QD_1F}(f). At the dip position, partial cancellation of spectral contributions suppresses the base response and reduces amplification, whereas at the peak position the emitter contribution dominates over the thermally active energy window, leading to enhanced gain. Despite the emergence of amplification, its magnitude remains moderate and varies smoothly with flux, reflecting the absence of inelastic scattering processes that could produce stronger spectral asymmetry and larger transistor gain.

\subsection{Voltage Probe: Engineered Reservoir and Linear-Response Enhancement}
We now consider the voltage-probe configuration of the first-generation SPG lattice [see Fig. \ref{fig:3QD}(a)], where the base terminal satisfies the condition $I_B=0$ while allowing a finite heat current $Q_B\neq0$. The self-consistent determination of the base chemical potential $\mu_B(\phi)$ [Eqs. \ref{vp_condition} and \ref{nr_method}] effectively converts the base into an inelastic reservoir, introducing feedback between the chemical potential and the heat currents and thereby modifying the transport properties relative to the purely fermionic case.\\
\indent
In the linear-response regime, Fig. \ref{fig:QD_1LV}(a)  shows that the amplification factor remains below unity throughout the investigated flux range. The corresponding emitter and base responses in [Fig. \ref{fig:QD_1LV}(b)] remain comparable. Thus, for the parameters considered here, the voltage-probe configuration does not produce amplification in the linear-response regime.
\section{Appendix D: Convergence Analysis}
Figure \ref{fig:convergence} represents the self-consistency convergence of the voltage-probe (base) chemical potential $(\mu_B)$, computed on SPG fractals of first-, second-, and third-generation, together with the amplification $(\alpha_{\mathrm{nl}})$. $\mu_B$ versus iteration number n for the first-, second-, and third-generation SPG at two flux values $\phi_d$ and $\phi_p$ are shown in Fig. \ref{fig:convergence}(a-c). In all three generations, $\mu_B$ rises sharply within the first $\sim$ 2-3 iterations and then saturates to a constant value. This behavior confirms the rapid and  stable convergence of the iterative scheme used to enforce zero particle current ($I_B=0$), at the voltage probe.\\
\indent
Lower panel Fig.\ref{fig:convergence}(a1-c1) shows the nonlinear amplification  ($\alpha_{\mathrm{nl}}$) plotted as a function of flux $\phi$. The close overlap between the two curves confirms that the results are well converged with respect to $d\omega$.\\
\indent
All numerical calculations presented in this work are performed using an energy discretization of $d\omega=10^{-6}$. The magnetic flux is varied uniformly over the range $-3\pi\leq\phi\leq3\pi$ using 307 equally spaced grid points.
\section{Appendix E : Physical Units}\label{appendixD}
We perform our simulations using the natural unit convention $\hbar=c=k_B=1$. It is very important to express our results in terms of physical units \cite{josefsson2018quantum}. Physical units can be obtained using the relation $\hbar\gamma_0=k_BT_a$, where $\gamma_0$ represents the dot-lead coupling and $T_a$ denotes the average temperature. Taking $T_a=1K$, we obtain $\gamma_0=1.3\times 10^{11}$ Hz. Taking the value of $\gamma_0$, we convert all parameters into physical units. In the simulation, we use the dot-lead coupling strength $\gamma=0.01$, which is converted to $\gamma=1.3$ GHz. $\gamma_0$ in terms of energy units is given by $\gamma_0=86.1$$\mu$eV. The parameters $\epsilon_d=20\gamma$, $T_E=60\gamma$, $T_C=10\gamma$, $\mu_E=-\mu_C=30\gamma$ translate to $\epsilon_d=17.2$$\mu$eV, $T_E=0.6$K, $T_C=0.1$K and $\mu_E=-\mu_C=25.8$$\mu$eV, respectively.

\bibliography{reff}

@article{li2006negative,
  title={Negative differential thermal resistance and thermal transistor},
  author={Li, Baowen and Wang, Lei and Casati, Giulio},
  journal={Applied Physics Letters},
  volume={88},
  number={14},
  year={2006},
  publisher={AIP Publishing}
}

@article{wang2007thermal,
  title={Thermal logic gates: computation with phonons},
  author={Wang, Lei and Li, Baowen},
  journal={Physical review letters},
  volume={99},
  number={17},
  pages={177208},
  year={2007},
  publisher={APS}
}

@article{li2012colloquium,
  title={Colloquium: Phononics: Manipulating heat flow with electronic<? format?> analogs and beyond},
  author={Li, Nianbei and Ren, Jie and Wang, Lei and Zhang, Gang and H{\"a}nggi, Peter and Li, Baowen},
  journal={Reviews of Modern Physics},
  volume={84},
  number={3},
  pages={1045--1066},
  year={2012},
  publisher={APS}
}

@article{segal2005spin,
  title={Spin-boson thermal rectifier},
  author={Segal, Dvira and Nitzan, Abraham},
  journal={Physical review letters},
  volume={94},
  number={3},
  pages={034301},
  year={2005},
  publisher={APS}
}

@article{benenti2017fundamental,
  title={Fundamental aspects of steady-state conversion of heat to work at the nanoscale},
  author={Benenti, Giuliano and Casati, Giulio and Saito, Keiji and Whitney, Robert S},
  journal={Physics Reports},
  volume={694},
  pages={1--124},
  year={2017},
  publisher={Elsevier}
}

@article{whitney2014most,
  title={Most efficient quantum thermoelectric at finite power output},
  author={Whitney, Robert S},
  journal={Physical review letters},
  volume={112},
  number={13},
  pages={130601},
  year={2014},
  publisher={APS}
}

@book{goldsmid2010introduction,
  title={Introduction to thermoelectricity},
  author={Goldsmid, H Julian and others},
  volume={121},
  year={2010},
  publisher={Springer}
}

@misc{imry1998introduction,
  title={Introduction to mesoscopic physics},
  author={Imry, Yoseph and Tinkham, Michael},
  year={1998},
  publisher={American Institute of Physics}
}

@book{datta1997electronic,
  title={Electronic transport in mesoscopic systems},
  author={Datta, Supriyo},
  year={1997},
  publisher={Cambridge university press}
}

@article{aharonov1959significance,
  title={Significance of electromagnetic potentials in the quantum theory},
  author={Aharonov, Yakir and Bohm, David},
  journal={Physical review},
  volume={115},
  number={3},
  pages={485},
  year={1959},
  publisher={APS}
}

@article{kobayashi2002tuning,
  title={Tuning of the Fano effect through a quantum dot in an Aharonov-Bohm interferometer},
  author={Kobayashi, Kensuke and Aikawa, Hisashi and Katsumoto, Shingo and Iye, Yasuhiro},
  journal={Physical review letters},
  volume={88},
  number={25},
  pages={256806},
  year={2002},
  publisher={APS}
}

@article{buttiker1988coherent,
  title={Coherent and sequential tunneling in series barriers},
  author={Buttiker, Markus},
  journal={IBM Journal of Research and Development},
  volume={32},
  number={1},
  pages={63--75},
  year={1988},
  publisher={IBM}
}

@article{bedkihal2013flux,
  title={Flux-dependent occupations and occupation difference in geometrically symmetric and energy degenerate double-dot Aharonov-Bohm interferometers},
  author={Bedkihal, Salil and Bandyopadhyay, Malay and Segal, Dvira},
  journal={Physical Review B—Condensed Matter and Materials Physics},
  volume={87},
  number={4},
  pages={045418},
  year={2013},
  publisher={APS}
}

@article{bedkihal2013magnetic,
  title={Magnetic field symmetries of nonlinear transport with elastic and inelastic scattering},
  author={Bedkihal, Salil and Bandyopadhyay, Malay and Segal, Dvira},
  journal={Physical Review B—Condensed Matter and Materials Physics},
  volume={88},
  number={15},
  pages={155407},
  year={2013},
  publisher={APS}
}

@article{bandyopadhyay2021flux,
  title={Flux dependent current rectification in geometrically symmetric interconnected triple-dot aharanov-bohm interferometer},
  author={Bandyopadhyay, Malay and Ghosh, Soumik and Dubey, A and Bedkihal, S},
  journal={Physica E: Low-dimensional Systems and Nanostructures},
  volume={133},
  pages={114786},
  year={2021},
  publisher={Elsevier}
}

@article{behera2023quantum,
  title={Quantum coherent control of nonlinear thermoelectric transport in a triple-dot Aharonov-Bohm heat engine},
  author={Behera, Jayasmita and Bedkihal, Salil and Agarwalla, Bijay Kumar and Bandyopadhyay, Malay},
  journal={Physical Review B},
  volume={108},
  number={16},
  pages={165419},
  year={2023},
  publisher={APS}
}

@article{bedkihal2025fundamental,
  title={Fundamental aspects of Aharonov--Bohm quantum machines: thermoelectric heat engines and diodes},
  author={Bedkihal, Salil and Behera, Jayasmita and Bandyopadhyay, Malay},
  journal={Journal of Physics: Condensed Matter},
  volume={37},
  number={16},
  pages={163001},
  year={2025},
  publisher={IOP Publishing}
}

@article{sridhar2026coherent,
  title={Coherent control of thermoelectric performance via engineered transmission functions in multidot Aharonov-Bohm heat engines},
  author={Sridhar and Bedkihal, Salil and Bandyopadhyay, Malay},
  journal={Physical Review B},
  volume={113},
  number={8},
  pages={085428},
  year={2026},
  publisher={APS}
}

@article{onsager1931reciprocal,
  title={Reciprocal relations in irreversible processes. I.},
  author={Onsager, Lars},
  journal={Physical review},
  volume={37},
  number={4},
  pages={405},
  year={1931},
  publisher={APS}
}

@misc{callen1998thermodynamics,
  title={Thermodynamics and an Introduction to Thermostatistics},
  author={Callen, Herbert B and Scott, HL},
  year={1998},
  publisher={American Association of Physics Teachers}
}

@article{benenti2011thermodynamic,
  title={Thermodynamic bounds on efficiency for systems with broken time-reversal symmetry},
  author={Benenti, Giuliano and Saito, Keiji and Casati, Giulio},
  journal={Physical review letters},
  volume={106},
  number={23},
  pages={230602},
  year={2011},
  publisher={APS}
}

@article{saito2011thermopower,
  title={Thermopower with broken time-reversal symmetry},
  author={Saito, Keiji and Benenti, Giuliano and Casati, Giulio and Prosen, Toma{\v{z}}},
  journal={Physical Review B—Condensed Matter and Materials Physics},
  volume={84},
  number={20},
  pages={201306},
  year={2011},
  publisher={APS}
}

@article{buttiker1986role,
  title={Role of quantum coherence in series resistors},
  author={B{\"u}ttiker, M},
  journal={Physical Review B},
  volume={33},
  number={5},
  pages={3020},
  year={1986},
  publisher={APS}
}

@article{buttiker1986four,
  title={Four-terminal phase-coherent conductance},
  author={B{\"u}ttiker, Markus},
  journal={Physical review letters},
  volume={57},
  number={14},
  pages={1761},
  year={1986},
  publisher={APS}
}

@article{christen1996gauge,
  title={Gauge-invariant nonlinear electric transport in mesoscopic conductors},
  author={Christen, Thomas and B{\"u}ttiker, M},
  journal={EPL (Europhysics Letters)},
  volume={35},
  number={7},
  pages={523--528},
  year={1996}
}

@article{sanchez2004magnetic,
  title={Magnetic-field asymmetry of nonlinear mesoscopic transport},
  author={S{\'a}nchez, David and B{\"u}ttiker, Markus},
  journal={Physical review letters},
  volume={93},
  number={10},
  pages={106802},
  year={2004},
  publisher={APS}
}

@article{meair2013scattering,
  title={Scattering theory of nonlinear thermoelectricity in quantum coherent conductors},
  author={Meair, Jonathan and Jacquod, Philippe},
  journal={Journal of Physics: Condensed Matter},
  volume={25},
  number={8},
  pages={082201},
  year={2013},
  publisher={IOP Publishing}
}

@book{haug2008quantum,
  title={Quantum kinetics in transport and optics of semiconductors},
  author={Haug, Hartmut and Jauho, Antti-Pekka},
  year={2008},
  publisher={Springer}
}

@article{bedkihal2013probe,
  title={The probe technique far from equilibrium: Magnetic field symmetries of nonlinear transport},
  author={Bedkihal, Salil and Bandyopadhyay, Malay and Segal, Dvira},
  journal={The European Physical Journal B},
  volume={86},
  number={12},
  pages={506},
  year={2013},
  publisher={Springer}
}

@article{AB1,
  title = {Manipulability of the Kondo effect in a T-shaped triple-quantum-dot structure},
  author = {Yi, Guang-Yu and Jiang, Cui and Zhang, Lian-Lian and Zhong, Su-Rui and Chu, Hao and Gong, Wei-Jiang},
  journal = {Phys. Rev. B},
  volume = {102},
  issue = {8},
  pages = {085418},
  numpages = {10},
  year = {2020},
  month = {Aug},
  publisher = {American Physical Society},
  doi = {10.1103/PhysRevB.102.085418},
  url = {https://link.aps.org/doi/10.1103/PhysRevB.102.085418}
}

@article{AB2,
  title = {Transport properties in a non-Hermitian triple-quantum-dot structure},
  author = {Zhang, Lian-Lian and Gong, Wei-Jiang},
  journal = {Phys. Rev. A},
  volume = {95},
  issue = {6},
  pages = {062123},
  numpages = {8},
  year = {2017},
  month = {Jun},
  publisher = {American Physical Society},
  doi = {10.1103/PhysRevA.95.062123},
  url = {https://link.aps.org/doi/10.1103/PhysRevA.95.062123}
}

@article{AB3,
  title={Josephson effect in a triple-quantum-dot ring with one dot coupled to superconductors: Numerical renormalization group calculations},
  author={Yi, Guang-Yu and Wang, Xiao-Qi and Gong, Wei-Jiang and Wu, Hai-Na and Chen, Xiao-Hui},
  journal={Physics Letters A},
  volume={380},
  number={14-15},
  pages={1385--1391},
  year={2016},
  publisher={Elsevier}
}

@article{AB4,
title = {Nonlocal magnetic configuration controlling realized in a triple-quantum-dot Josephson junction},
journal = {Physica E: Low-dimensional Systems and Nanostructures},
volume = {81},
pages = {26-30},
year = {2016},
issn = {1386-9477},
doi = {https://doi.org/10.1016/j.physe.2016.02.013},
url = {https://www.sciencedirect.com/science/article/pii/S138694771630056X},
author = {Guang-Yu Yi and Xiao-Qi Wang and Hai-Na Wu and Wei-Jiang Gong}
}

@article{dharsen,
  title = {Nonequilibrium Green's function formalism and the problem of bound states},
  author = {Dhar, Abhishek and Sen, Diptiman},
  journal = {Phys. Rev. B},
  volume = {73},
  issue = {8},
  pages = {085119},
  numpages = {14},
  year = {2006},
  month = {Feb},
  publisher = {American Physical Society},
  doi = {10.1103/PhysRevB.73.085119},
  url = {https://link.aps.org/doi/10.1103/PhysRevB.73.085119}
}

@article{pooja,
  title = {Landauer formula for the current through an interacting electron region},
  author = {Meir, Yigal and Wingreen, Ned S.},
  journal = {Phys. Rev. Lett.},
  volume = {68},
  issue = {16},
  pages = {2512--2515},
  numpages = {0},
  year = {1992},
  month = {Apr},
  publisher = {American Physical Society},
  doi = {10.1103/PhysRevLett.68.2512},
  url = {https://link.aps.org/doi/10.1103/PhysRevLett.68.2512}
}

@article{wang2014nonequilibrium,
  title={Nonequilibrium Green’s function method for quantum thermal transport},
  author={Wang, Jian-Sheng and Agarwalla, Bijay Kumar and Li, Huanan and Thingna, Juzar},
  journal={Frontiers of Physics},
  volume={9},
  pages={673--697},
  year={2014},
  publisher={Springer}
}

@book{fransson2010non,
  title={Non-equilibrium nano-physics: a many-body approach},
  author={Fransson, Jonas},
  volume={809},
  year={2010},
  publisher={Springer}
}

@article{PhysRevB.92.045309,
  title = {Phonon thermoelectric transistors and rectifiers},
  author = {Jiang, Jian-Hua and Kulkarni, Manas and Segal, Dvira and Imry, Yoseph},
  journal = {Phys. Rev. B},
  volume = {92},
  issue = {4},
  pages = {045309},
  numpages = {9},
  year = {2015},
  month = {Jul},
  publisher = {American Physical Society},
  doi = {10.1103/PhysRevB.92.045309},
  url = {https://link.aps.org/doi/10.1103/PhysRevB.92.045309}
}

@article{PhysRevB.87.205420,
  title = {Hopping thermoelectric transport in finite systems: Boundary effects},
  author = {Jiang, Jian-Hua and Entin-Wohlman, Ora and Imry, Yoseph},
  journal = {Phys. Rev. B},
  volume = {87},
  issue = {20},
  pages = {205420},
  numpages = {12},
  year = {2013},
  month = {May},
  publisher = {American Physical Society},
  doi = {10.1103/PhysRevB.87.205420},
  url = {https://link.aps.org/doi/10.1103/PhysRevB.87.205420}
}

@article{PhysRevB.85.085401,
  title = {Three-terminal thermoelectric transport under broken time-reversal symmetry},
  author = {Entin-Wohlman, O. and Aharony, A.},
  journal = {Phys. Rev. B},
  volume = {85},
  issue = {8},
  pages = {085401},
  numpages = {10},
  year = {2012},
  month = {Feb},
  publisher = {American Physical Society},
  doi = {10.1103/PhysRevB.85.085401},
  url = {https://link.aps.org/doi/10.1103/PhysRevB.85.085401}
}

@article{pal2025fractal,
  title={A fractal geometry immersed in a hierarchical magnetic flux distribution},
  author={Pal, Biplab},
  journal={Journal of Applied Physics},
  volume={137},
  number={14},
  year={2025},
  publisher={AIP Publishing}
}

@article{PhysRevB.56.13768,
  title = {Sierpi{\'n}ski gasket in a magnetic field: Electron states and transmission characteristics},
  author = {Chakrabarti, Arunava and Bhattacharyya, Bibhas},
  journal = {Phys. Rev. B},
  volume = {56},
  issue = {21},
  pages = {13768--13773},
  numpages = {0},
  year = {1997},
  month = {Dec},
  publisher = {American Physical Society},
  doi = {10.1103/PhysRevB.56.13768},
  url = {https://link.aps.org/doi/10.1103/PhysRevB.56.13768}
}

@article{PhysRevB.97.195101,
  title = {Flat bands in fractal-like geometry},
  author = {Pal, Biplab and Saha, Kush},
  journal = {Phys. Rev. B},
  volume = {97},
  issue = {19},
  pages = {195101},
  numpages = {7},
  year = {2018},
  month = {May},
  publisher = {American Physical Society},
  doi = {10.1103/PhysRevB.97.195101},
  url = {https://link.aps.org/doi/10.1103/PhysRevB.97.195101}
}

@article{josefsson2018quantum,
  title={A quantum-dot heat engine operating close to the thermodynamic efficiency limits},
  author={Josefsson, Martin and Svilans, Artis and Burke, Adam M and Hoffmann, Eric A and Fahlvik, Sofia and Thelander, Claes and Leijnse, Martin and Linke, Heiner},
  journal={Nature nanotechnology},
  volume={13},
  number={10},
  pages={920--924},
  year={2018},
  publisher={Nature Publishing Group UK London}
}

@article{PhysRevB.51.9310,
  title = {Localization in fractal spaces: Exact results on the Sierpi{\'n}ski gasket},
  author = {Wang, Xiang Rong},
  journal = {Phys. Rev. B},
  volume = {51},
  issue = {14},
  pages = {9310--9313},
  numpages = {0},
  year = {1995},
  month = {Apr},
  publisher = {American Physical Society},
  doi = {10.1103/PhysRevB.51.9310},
  url = {https://link.aps.org/doi/10.1103/PhysRevB.51.9310}
}

@article{Jana2010,
  author  = {Supriya Jana and Arunava Chakrabarti and Samar Chattopadhyay},
  title   = {Electronic transport in an anisotropic Sierpi{\'n}ski gasket},
  journal = {Physica B: Condensed Matter},
  volume  = {405},
  number  = {17},
  pages   = {3735--3740},
  year    = {2010},
  doi     = {10.1016/j.physb.2010.05.077},
  issn    = {0921-4526},
  publisher = {Elsevier}
}

@article{Landauer1970,
  author    = {Rolf Landauer},
  title     = {Electrical Resistance of Disordered One-Dimensional Lattices},
  journal   = {Philosophical Magazine},
  volume    = {21},
  number    = {172},
  pages     = {863--867},
  year      = {1970},
  doi       = {10.1080/14786437008238472}
}

@article{Gorbatsevich2018PT,
  author  = {Gorbatsevich, A. A. and Krasnikov, G. Ya. and Shubin, N. M.},
  title   = {$\mathcal{P}\mathcal{T}$-symmetric interference transistor},
  journal = {Scientific Reports},
  volume  = {8},
  pages   = {15780},
  year    = {2018},
  doi     = {10.1038/s41598-018-34132-0},
  url     = {https://doi.org/10.1038/s41598-018-34132-0}
}
\end{document}